\documentclass[aps,pre,groupedaddress]{revtex4}

\usepackage{amsmath}

\usepackage{amssymb}

\usepackage{amsbsy}

\usepackage{graphicx}

\usepackage{color}

\begin{document}

\title{Nonlinear gap modes and compactons in a lattice
  model for spin-orbit coupled exciton-polaritons in zigzag chains}

\author{Magnus Johansson}
\email{mjn@ifm.liu.se}
\homepage{https://people.ifm.liu.se/majoh}
\affiliation{Department of Physics, Chemistry and Biology, Link\"oping University, SE-581 83 Link\"oping, Sweden}
\author{Petra P.\ Beli\v cev}
\affiliation{P$^{*}$ Group, Vin\v ca Institute of Nuclear
Sciences, University of Belgrade, P.O.\ Box 522, 11001 Belgrade, Serbia}
\author{Goran Gligori\'c}
\affiliation{P$^{*}$ Group, Vin\v ca Institute of Nuclear
Sciences, University of Belgrade, P.O.\ Box 522, 11001 Belgrade, Serbia}
\author{Dmitry R.\ Gulevich}
\affiliation{ITMO University, St.\ Petersburg 197101, Russia}
\author{Dmitry V.\ Skryabin}
\affiliation{Department of Physics, University of Bath, Bath BA2 7AY,
United Kingdom}
\affiliation{ITMO University, St.\ Petersburg 197101, Russia}

\begin{abstract}
We consider a system of
generalized coupled Discrete Nonlinear Schr{\"o}dinger (DNLS) equations,
derived as a tight-binding model from the
Gross-Pitaevskii-type equations describing a zigzag chain of weakly coupled
condensates of exciton-polaritons with
spin-orbit (TE-TM) coupling. We focus on the simplest case when the
angles for the links in the zigzag chain
are $\pm \pi/4$ with respect to the chain axis, 
and the basis (Wannier) functions are cylindrically symmetric
(zero orbital angular momenta). We analyze the properties
of the fundamental nonlinear localized solutions, with particular interest
in the discrete gap solitons appearing due to the simultaneous presence of
spin-orbit coupling and zigzag geometry, opening a gap in the linear dispersion
relation. In particular, their linear stability is analyzed. We also find
that the linear dispersion relation becomes exactly flat at particular
parameter values, and obtain corresponding compact solutions
localized on two neighboring sites, with spin-up and spin-down parts
$\pi/2$ out of phase at each site. The continuation of these compact modes
into exponentially decaying gap modes for generic parameter values is studied
numerically, and regions of stability are found to exist in the lower or upper half of
the gap, depending on the type of gap modes.
\end{abstract}

\date{\today}

\pacs{} \maketitle

\section{Introduction}

Planar semiconductor microcavities operating in the exciton-polariton regime  
have become a paradigm model for experimental and theoretical studies of 
nonlinear and quantum properties of light-matter  interaction \cite{ds1}. 
A major advantage of these systems is that they are solid state devices, that 
operate in a wide diapason of temperatures between few Kelvins and up to the 
room conditions. Interaction between the polaritons is much stronger than for 
pure photons, that lowers power requirements for creating conditions when 
polariton dynamics can be effectively controlled with the external light 
sources \cite{ds2}. Microcavities can also be readily structured to create a 
variety of potential energy landscapes reproducing lattice structures known in 
studies of electrons in condensed matter on more practical scales of tens of 
microns. Thus polaritons can be controlled using band gap and zone 
engineering \cite{ds3}. 
Through their peculiar spin properties and sensitivity to 
the applied magnetic field, polaritons in structured microcavities have been 
shown to have a number of topological properties \cite{ds4}. 
Thus polariton based 
devices have a competitive edge over their photon-only counterparts through 
their relatively low nonlinear thresholds and possibility to create 
micron-scale topological devices. A combination of these two aspects has been 
recently used to demonstrate a variety of nonlinear topological effects in 
polariton systems, see, e.g., \cite{ds5} and references therein.

As a specific example, a polariton BEC in a zigzag chain of polariton 
micropillars with photonic spin-orbit coupling, originating in the splitting 
of optical cavity modes with TE and TM polarization, was proposed in 
Ref.\ \cite{Solnyshkov16}. The simultaneous presence of zigzag geometry and 
polarization dependent tunneling was shown to yield topologically protected 
edge states, and in the presence of homogeneous pumping and nonlinear 
interactions the creation of polarization domain walls through the Kibble-Zurek
mechanism, analogous to the Su-Schrieffer-Heeger solitons in polymers, was 
numerically observed \cite{Solnyshkov16}. Of crucial importance is the 
spin-orbit induced opening of a central gap in the linear dispersion relation.
As we will show in this work, the existence of a gap, together with the option 
of tuning the linear dispersion towards flatness at specific parameter values, 
also leads to nonlinear strongly localized modes in the bulk 
(intrinsically localized modes) with properties depending crucially on the 
relative strength of interaction between polaritons of opposite and equal spin.


The starting point is the following set of two coupled continuous
Gross-Pitaevskii equations \cite{Flayac10}:
\begin{eqnarray}
  i \partial_t{\Psi}_+ = - \frac{1}{2}(\partial_x^2 + \partial_y^2) \Psi_+
  + \left(|\Psi_+|^2 + \mathfrak{a} |\Psi_-|^2\right) \Psi_+ + \Omega \Psi_+
  + \beta(\partial_x -i \partial_y)^2 \Psi_-+ V(x,y){\Psi}_+
 \nonumber  \\
  i \partial_t{\Psi}_- = - \frac{1}{2}(\partial_x^2 + \partial_y^2) \Psi_-
  + \left(|\Psi_-|^2 + \mathfrak{a} |\Psi_+|^2\right) \Psi_- - \Omega \Psi_-
  + \beta(\partial_x +i \partial_y)^2\Psi_+ + V(x,y){\Psi}_- .
 \label{cont}
 \end{eqnarray}
These equations describe exciton-polaritons with circularly polarized
light-component, where $\Psi_+$ corresponds to left (positive spin) and
$\Psi_-$ to right (negative spin)
polarization. Polaritons interact mainly through their excitonic part, and
interactions between polaritons with identical polarization are generally
repulsive (here normalized to +1), while interactions between those of opposite
spins often are weaker and attractive. A typical value
is $\mathfrak{a} \simeq -0.05$ \cite{Sich14}, but may range between roughly
$-1 \lesssim \mathfrak{a} \lesssim 0$, and may possibly be also repulsive, or
attractive with a magnitude stronger than the self-interaction
\cite{Vladimirova10}. Since the exciton-components of the polariton
wave functions typically are localized within small spatial regions, the
interactions are assumed to be local (point interactions) in this mean-field
description. $\Omega$ describes the Zeeman-splitting between spin-up and
spin-down polaritons in presence of an external magnetic field; in this
work we put $\Omega=0$.

Of main interest here is the term proportional to $\beta$: it arises due to
different properties associated with polaritons whose photonic components,
as expressed in a suitable basis of linear polarization, have TE resp TM
polarizations (or, alternatively, longitudinal/transversal w.r.t. the
propagation direction (${\bf k}$-vector)).
It is commonly described in terms of different effective masses of the
lower polariton branches
for TE and TM components, $\beta \propto m_{TE}^{-1}-m_{TM}^{-1}$, whose ratio
typically may be of the order $m_{TE}/m_{TM} \approx 0.85-0.95$ (see e.g.
supplemental material of \cite{Dufferwiel15}), although in principle
$\beta$ could have arbitrary sign. Expressed in a basis of circular
polarization (spinor basis) as in (\ref{cont})
($\Psi_{\pm}=\Psi_x \mp i \Psi_y$),
this TE/TM energy splitting can be interpreted as a spin-orbit splitting,
since the dynamics of the two spin (polarization) components couple in
a different way to the orbital part of the other component (via derivatives
in $x$ and $y$ of the mean-field wave function in (\ref{cont})).

\begin{figure}
\includegraphics[height=0.49\textwidth,angle=270]{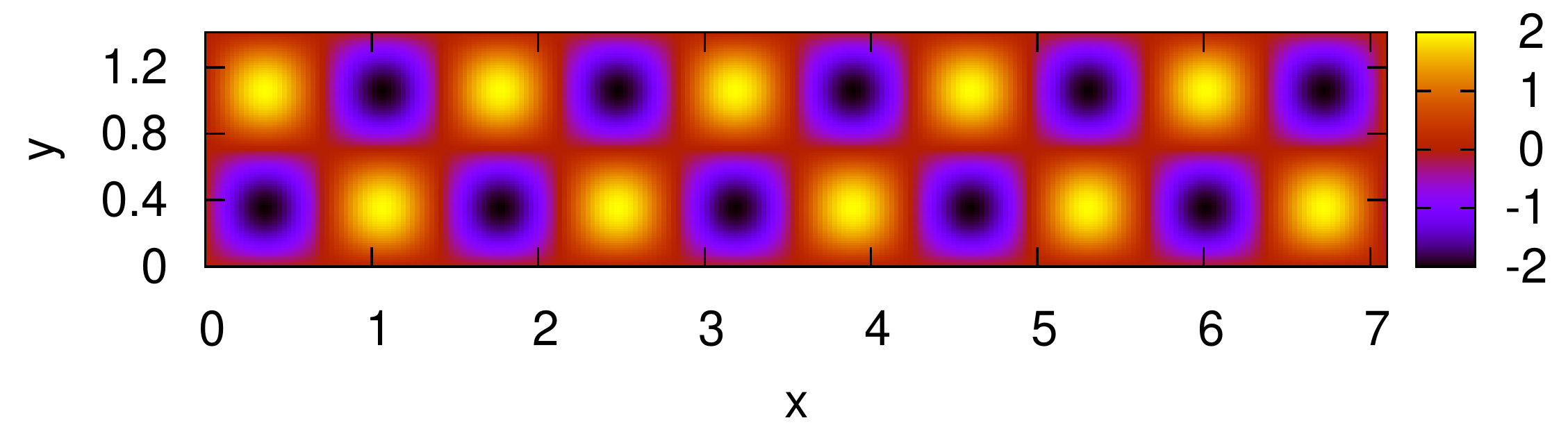}
\caption{The zigzag potential $V(x,y)$ (\ref{pot}) with $V_0=d=1$ and $N=5$. 
In the tight-binding expansion (\ref{expansion}), the Wannier functions 
are assumed to be centered around the lattice minima.
}
\label{fig:zig-zag}
\end{figure}

In this work, we choose the potential $V(x,y)$ as a zigzag potential along
the $x$-direction, considering this geometry as the simplest generalization of
a straight 1D chain which yields non-trivial geometrical
effects of the spin-orbit coupling between
polaritons localized at neighboring potential minima. As an
example potential, we may choose e.g.:
\begin{equation}
  V(x,y) = - 2V_0 \sin \left(\frac{\sqrt{2} \pi}{d} x\right)
  \sin{\left(\frac{\sqrt{2} \pi}{d} y\right)}; \quad
  0\leq x \leq N\sqrt{2} d, \ 0 \leq y \leq \sqrt{2} d,
\label{pot}
\end{equation}
as illustrated in Fig.\ \ref{fig:zig-zag}. Here $d$ is the distance between
potential mininma, and $2N$ is the total number of potential wells in the
chain. The geometry is essentially the same as for the coupled micropillars
in \cite{Solnyshkov16}, with all angles for the links between neighboring 
minima being $\pm 45^\circ$ with respect to the $x$-axis. Evidently one may
easily generalize to arbitrary angles, or more complicated
expressions for zigzag potentials which may be realized in various
experimental settings e.g. with optical lattices \cite{Zhang15}. In order to
motivate a tight-binding approximation, we assume $V_0 \gg 1$.

In order to understand the most important effects of the spin-coupling
coupling in
(\ref{cont}) in a tight-binding framework, we here consider situations where
the effects of spin-orbit splitting {\em inside} each potential well can be
neglected, and only are relevant in the regions of wavefunction overlap
{\em between} neighboring wells. For the experimental set-up of
\cite{Dufferwiel15}, this should be a good approximation if the spatial modes
inside the wells may be approximated with Laguerre-Gauss modes with zero
orbital angular momentum ($LG_{00}^\pm$ in the notation of \cite{Dufferwiel15},
where the two subscripts stand for radial and orbital quantum numbers
of the 2D harmonic-oscillator wave function,
and the superscript indicates polarization as in (\ref{cont}).) At least
for a single cavity of non-interacting polaritons, these modes should be
good approximations to the ground state, so let us assume that interactions
(nonlinearity) and spin-orbit couplings are sufficiently weak to be treated
perturbatively, along with the inter-well overlaps. The approach may be
extended to
consider also lattices of spin vortices (excited modes)
built up from modes with nonzero OAM
(e.g. $LG_{0\pm 1}^\pm$ as considered in \cite{Dufferwiel15}); however this will
introduce some additional complications and will be left for future work.

Moreover, if $V_0 \gg 1$ we may also neglect the effect of
next-nearest-neighbor interactions (distances between two wells
in the horizontal $x$-direction is $\sqrt{2}$ times larger than between
nearest neighbors). It may then be a good approximation to use, as
the basis set for the tight-binding approximation, the Wannier functions for
a full 2D square lattice (these issues are discussed and numerically
checked for some realization of a zigzag optical lattice in a recent Master
thesis \cite{Gagge16}). These may resemble (but certainly differ from)
\cite{Gagge16} the LG individual modes (e.g. Wannier functions typically have
radial oscillatory tails, decaying exponentially rather than Gaussian).
In any case, we will assume that the basis
functions $w(x,y)$ (expressed in Cartesian coordinates)
are qualitatively close to the $LG_{00}$ modes. Particularly, they will be
assumed to be close to cylindrically symmetric
($w(x,y)\sim e^{-\omega(x^2+y^2)}$ in the harmonic approximation).
(Note that this assumption would not be valid for spin vortices arising
from LG modes with nonzero OAM.)

The outline of this paper is as follows. In Sec.\ \ref{sec:model} we 
derive the tight-binding model, discuss its general properties, and 
illustrate the linear dispersion relation for the case with $\pm 45^\circ$
angles which will be the system studied for the rest of this paper. 
We also in Sec.\ \ref{sec:flat} identify a limit where the linear dispersion 
relation becomes exactly flat, and identify the corresponding fundamental 
compact solutions. In Sec.\ \ref{sec:nonlinear} we construct the fundamental 
nonlinear localized modes in the semi-infinite gaps above or below the linear 
spectrum, as well as in the mini-gap between the linear dispersion branches, 
opened up due to the simultaneous presence of spin-orbit coupling and 
nontrivial geometry. Analytical calculations using perturbation theory 
from the weak-coupling and flat-band limits for the semi-infinite and mini-gap, 
respectively, are compared with numerical calculations using a standard 
Newton scheme.  In Sec.\ \ref{sec:stab} the linear stability of the 
different families of nonlinear localized modes is investigated, and some 
instability scenarios are illustrated with direct dynamical simulations. 
Finally, some concluding remarks are given in Sec.\ \ref{sec:conc}.

\section{Model}
\label{sec:model}
\subsection{Derivation of the tight-binding model}
Under the above assumptions, we may expand:
\begin{equation}
  \Psi_+ = \sum_{n=1}^{2N}u_n(t) w(x'-nd', y'-(-1)^n d'/2), \quad
  \Psi_- = \sum_{n=1}^{2N}v_n(t) w(x'-nd', y'-(-1)^n d'/2),
  \label{expansion}
\end{equation}
where, relative to the coordinate system of (\ref{pot}) 
and Fig.\ \ref{fig:zig-zag},
$d'=d/\sqrt{2}, x'=x-d'/2, y'=y-d'$. Note that the (Wannier) basis functions
are the same for both components, since we have assumed no spin-orbit splitting
inside the wells, $\Omega=0$, and $w$ are basis functions of the linear
problem. Note also that an analogous approach was used in
\cite{Salerno15} to derive lattice equations for the simpler problem of a
pure 1D lattice with a standard spin-orbit coupling term ($-i \partial_x$,
linear in the spatial derivative) for atomic BEC's in optical lattices;
similar models were also studied in \cite{Sakaguchi14,Belicev15,Gligoric16}.
For simplicity we will
assume below that $w(x,y)$
can be chosen real (which is typically the case in absence of OAM;
the generalization to modes with
nonzero OAM requires complex  $w(x,y)$ and will be treated
in a separate work).

Inserting the expansion (\ref{expansion}) into (\ref{cont}), we obtain for
the first component:

\begin{eqnarray}
  i \sum_{n'}\dot{u}_{n'} w(x'-n'd', y'-(-1)^{n'} d'/2)
  \nonumber\\
  =
  -\frac{1}{2} \sum_{n'}{u}_{n'}\left[w_{xx}(x'-n'd', y'-(-1)^{n'} d'/2) +
    w_{yy}(x'-n'd', y'-(-1)^{n'} d'/2)\right]
  \nonumber\\
  +\sum_{n'}\left(|u_{n'}|^2+\mathfrak{a}|v_{n'}|^2 \right)u_{n'}
  w^3(x'-n'd', y'-(-1)^{n'} d'/2)
  \nonumber\\
  +\beta \sum_{n'}{v}_{n'}\left[w_{xx}(x'-n'd', y'-(-1)^{n'} d'/2) -
    w_{yy}(x'-n'd', y'-(-1)^{n'} d'/2)-2i w_{xy}(x'-n'd', y'-(-1)^{n'} d'/2 \right]
  \nonumber\\
  +V(x,y) \sum_{n'}{u}_{n'} w(x'-n'd', y'-(-1)^{n'} d'/2) .
  \label{insert}
\end{eqnarray}
Here, in writing the nonlinear term as a simple sum and not a triple, we have
neglected overlap between basis functions on different sites in cubic terms 
in $w$ (assuming strong localization of $w$).

Multiplying with $w^{(n)}\equiv w(x'-nd', y'-(-1)^{n} d'/2)$,
integrating over $x$ and $y$,
using the orthogonality of Wannier functions and neglecting all overlaps
beyond nearest neighbors, we obtain from (\ref{insert}) a 1D lattice equation
of the following form for the site amplitudes of the spin-up component:
\begin{equation}
  i\dot{u}_{n} = \epsilon u_n - \Gamma \left(u_{n+1}+u_{n-1}\right)
  + \gamma \left(|u_{n}|^2+\mathfrak{a}|v_{n}|^2\right)u_n
  + \omega v_n + \sigma_{n,n+1} v_{n+1} + \sigma_{n,n-1} v_{n-1}.
  \label{u}
\end{equation}
Here the coefficients are:
On-site energy,
$$
\epsilon = \int\int\left[-\frac{1}{2}(w^{(n)}_{xx}+w^{(n)}_{yy}) +
  V(x,y) w^{(n)} \right] w^{(n)}dx dy;
$$
linear coupling coefficients,
$$
\Gamma = \frac{1}{2}\int\int(w^{(n+1)}_{xx}+w^{(n+1)}_{yy})w^{(n)}dx dy=
  \frac{1}{2}\int\int(w^{(n-1)}_{xx}+w^{(n-1)}_{yy}) w^{(n)}dx dy,
  $$
  where the second equality is obviously true if $w^{(n)}$ is cylindrically
  symmetric; nonlinearity coefficient,
  $$
  \gamma = \int\int  (w^{(n)})^4 dx dy;
  $$
  on-site spin-orbit interaction,
  $$
  \omega=\beta\int\int (w^{(n)}_{xx}-w^{(n)}_{yy}-2iw^{(n)}_{xy}) w^{(n)}dx dy,
  $$
  which is identically zero if $w^{(n)}$ is cylindrically symmetric
  (easiest seen in polar coordinates, with $w=w(r)$ only,
  $\omega=\beta\int_0^{2\pi}d\phi e^{-2i\phi}\int r dr (w_{rr}-\frac{w_r}{r})w=0$)
  (but generally nonzero if Wannier modes would have OAM);
  and nearest-neighbor spin-orbit interactions (the relevant 'new' terms here),
  \begin{equation}
  \sigma_{n,n\pm 1} = \beta \int\int
  (w^{(n\pm 1)}_{xx}-w^{(n\pm 1)}_{yy}-2iw^{(n\pm 1)}_{xy}) w^{(n)}dx dy.
  \label{sigmadef}
  \end{equation}

  Since tails of $w$ are exponentially small, we may assume all integrals
  taken over the infinite plane. Explicitly, with a change of origin we may
  write e.g. the first term in the integral in (\ref{sigmadef}) as
  $\int\int w_{xx}(x\mp d', y-(-1)^n d')w(x,y)dxdy$, etc. But for the case
  with $w$ cylindrically symmetric, we may easier evaluate the integral
  (\ref{sigmadef}) in polar coordinates, centered at site $n\pm 1$. After
  some elementary trigonometry we then obtain:
  \begin{equation}
    \sigma_{n,n\pm 1} = \beta \int\int e^{-i 2 \phi}\left(w_{rr}-\frac{w_r}{r}\right)
    w(\sqrt{d^2+r^2\pm 2dr\cos\left(\frac{\pi}{4}\pm (-1)^n\phi\right)})
    rdrd\phi .
    \label{trig}
  \end{equation}
  Letting $\phi' = \frac{\pi}{4} \pm (-1)^n \phi$, this can be expressed as
  \begin{equation}
    \sigma_{n,n\pm 1} = e^{\pm 2 i \alpha_n} \sigma;\quad
    \sigma \equiv \beta \int\int e^{\mp (-1)^n i 2 \phi'}
    \left(w_{rr}-\frac{w_r}{r}\right) w(\sqrt{d^2+r^2\pm 2dr\cos\phi'})
    rdrd\phi' ,
    \label{sigmadef2}
    \end{equation}
  where $\alpha_n \equiv (-1)^n\frac{\pi}{4}$ are the angles for the links in
  the zigzag chain with respect to the $x$-axis, and the integral defining
  $\sigma$ is independent of all signs since $\cos \phi'$ is even.
  Explicitly, for the $\pi/4$ zigzag chain we get
  \begin{equation}
    \sigma_{n,n\pm1} = \left\{\begin{array}{rl}
        -i\sigma & \text{diagonal links}
        \\
        +i\sigma & \text{antidiagonal links}
    \end{array}\right.
    \label{diag}.
  \end{equation}

  Proceeding analogously with the second component, we obtain the corresponding
  lattice equation for the site-amplitudes of the spin-down component:
  \begin{equation}
  i\dot{v}_{n} = \epsilon v_n - \Gamma \left(v_{n+1}+v_{n-1}\right)
  + \gamma \left(|v_{n}|^2+\mathfrak{a}|u_{n}|^2\right)v_n
  + \omega' u_n + \sigma'_{n,n+1} u_{n+1} + \sigma'_{n,n-1} u_{n-1}.
  \label{v}
\end{equation}
  Here, $\epsilon, \Gamma, \gamma$ are identical as for the first component
  (i.e., we may put $\epsilon=0$ by redefining zero-energy, and
  $\gamma=1$ (or alternatively $\Gamma=1$) by redefining energy scale). For
  the on-site spin-orbit interaction,
$$  \omega'=\beta\int\int (w^{(n)}_{xx}-w^{(n)}_{yy}+2iw^{(n)}_{xy}) w^{(n)}dx dy,
  $$
  (note opposite sign of third term compared to $\omega$), which is again
  zero if $w$ is cylindrically symmetric. And finally, for the
  nearest-neighbor spin-orbit couplings,
   \begin{equation}
  \sigma'_{n,n\pm 1} = \beta \int\int
  (w^{(n\pm 1)}_{xx}-w^{(n\pm 1)}_{yy}+2iw^{(n\pm 1)}_{xy}) w^{(n)}dx dy
  \label{sigmaprimedef}
  \end{equation}
   (again note sign of third term compared to (\ref{sigmadef}). As before,
   restricting to cylindrically symmetric $w$ yields
   \begin{eqnarray}
     \sigma'_{n,n\pm 1} = \beta \int\int e^{+i 2 \phi}\left(w_{rr}-\frac{w_r}{r}
     \right)
    w(\sqrt{d^2+r^2\pm 2dr\cos\left(\frac{\pi}{4}\pm (-1)^n\phi\right)})
    rdrd\phi
    \nonumber\\
    = \beta e^{\mp (-1)^n i\pi/2}\int\int e^{\pm (-1)^n i 2 \phi'}
    \left(w_{rr}-\frac{w_r}{r}\right) w(\sqrt{d^2+r^2\pm 2dr\cos\phi'})
    rdrd\phi'
    = e^{\mp 2 i \alpha_n} \sigma,
 \label{sigmaprime}    \end{eqnarray}
   where the last equality holds since the integral is equivalent to that of
   (\ref{sigmadef2}).  Explicitly, for the $\pi/4$ zigzag chain
  \begin{equation}
    \sigma'_{n,n\pm1} = \left\{\begin{array}{rl}
        +i\sigma & \text{diagonal links}
        \\
        -i\sigma & \text{antidiagonal links}
    \end{array}\right.
    \label{diag2},
  \end{equation}
  i.e., with opposite signs compared to (\ref{diag}). Note that, under the
  above assumptions ($w$ real and cylindrically symmetric), the integral
  defining $\sigma$ is always real.

  We note that the resulting lattice equations (\ref{u}), (\ref{v}),
  with spin-orbit coefficients  given by (\ref{diag}), (\ref{diag2}), are not
  equivalent to the equations studied in
  \cite{Salerno15,Belicev15,Sakaguchi14,Gligoric16}. In particular, we comment
  on the relation between the present model and that of Ref.\ \cite{Gligoric16},
  who considered a diamond chain with angles $\pi/4$ and a Rashba-type
  spin-orbit coupling. The zigzag chain could be considered as e.g.\
  the upper part of the diamond chain, if all amplitudes of the lower strand
  would vanish. However, because the spin-orbit coupling used in
  \cite{Gligoric16} is linear in the spatial derivatives while in this work
  it is  quadratic, the spin-orbit coupling coefficients in \cite{Gligoric16}
  have a phase shift of $\pi/2$ between diagonal and antidiagonal links,
  compared to $\pi$ in (\ref{diag}), (\ref{diag2}).

  \subsection{General properties of the TB-equations}
  \label{general}

  Let us put $\epsilon=0$ and $\gamma=1$. As above, assuming cylindrically
  symmetric basis functions, we have $\omega=\omega'=0$. We 
also remind the reader that we consider the case with no external
  magnetic field, $\Omega=0$ in (\ref{cont}).
 Equations (\ref{u}) and (\ref{v}),
 with spin-orbit coefficients  given by (\ref{diag}) and (\ref{diag2}),
 respectively, then become:
\begin{eqnarray}
  i\dot{u}_{n} = - \Gamma \left(u_{n+1}+u_{n-1}\right)
  + \left(|u_{n}|^2+\mathfrak{a}|v_{n}|^2\right)u_n
  + (-1)^n i \sigma \left(v_{n+1} -v_{n-1}\right) \quad \text{(spin-up)};
  \nonumber\\
 i\dot{v}_{n} = - \Gamma \left(v_{n+1}+v_{n-1}\right)
  + \left(|v_{n}|^2+\mathfrak{a}|u_{n}|^2\right)v_n
  - (-1)^n i \sigma \left(u_{n+1} -u_{n-1}\right) \quad \text{(spin-down)} .
  \label{uvdnls}
\end{eqnarray}

One may easily show the existence of the ``standard'' two conserved quantities
for DNLS-type models; Norm (Power):
\begin{equation}
  P=\sum_n\left(|u_n|^2 + |v_n|^2\right),
\end{equation}
  and Hamiltonian:
\begin{equation}
  H = \sum_n\left\{ -\Gamma \left(u_n^* u_{n+1} + v_n^* v_{n+1}\right)
  + \frac{1}{4}\left(|u_n|^4 + |v_n|^4\right) + \frac{\mathfrak{a}}{2}|u_n|^2|v_n|^2
  +(-1)^n i \sigma u_n^*\left(v_{n+1}-v_{n-1}\right)
  \right\} + c.c. .
  \label{Ham}
\end{equation}
Here, $\{u_n, v_n\}$ and $\{iu^*_n, iv^*_n\}$ play the role of conjugated
coordinates and momenta, respectively (i.e.,
$\dot{u}_n=\partial H/\partial(iu^*_n), \dot{v}_n=\partial H/\partial(iv^*_n)$,
etc.). We may note that the Hamiltonian is similar to the Hamiltonian for
the ``inter-SOC'' chain of Beli{\v c}ev et al. (Eq. (11) in \cite{Belicev15}),
but differs by the ``zigzag'' spin-orbit factor $(-1)^n$ in the last term.
Note that this factor can be removed by performing a
``staggering transformation'' on the site-amplitudes of the spin-down
component: $v_n' = (-1)^n v_n$, transforming the equations of motion
(\ref{uvdnls}) into:
\begin{eqnarray}
  i\dot{u}_{n} = - \Gamma \left(u_{n+1}+u_{n-1}\right)
  + \left(|u_{n}|^2+\mathfrak{a}|v_{n}'|^2\right)u_n
  - i \sigma \left(v_{n+1}' -v_{n-1}'\right) \quad \text{(spin-up)};
  \nonumber\\
 i\dot{v}_{n}' = + \Gamma \left(v_{n+1}'+v_{n-1}'\right)
  + \left(|v_{n}'|^2+\mathfrak{a}|u_{n}|^2\right)v_n'
  - i \sigma \left(u_{n+1} -u_{n-1}\right) \quad \text{(spin-down)} .
  \label{uvprime}
\end{eqnarray}
Thus, this transformation effectively changes the sign of the linear
coupling of the spin-down component into
$\Gamma \rightarrow \Gamma'=-\Gamma$ (which may be interpreted as a
reversal of the ``effective mass'' of the spin-down polariton in this
tight-binding approximation), while the nonlinear and spin-orbit terms
for both components become equivalent. Eqs. (\ref{uvprime}) differ from
the equations derived in \cite{Salerno15} for the straight chain
with standard spin-orbit coupling only through this sign-change
of $\Gamma$ for the spin-down component.
Note also that Eqs.\ (\ref{uvprime}) are invariant under a transformation
$v_{n}' \rightarrow - v_{n}'$, $n+1 \rightarrow n-1$, i.e., an overall
change of the
relative sign of the spin-up and spin-down components is equivalent to
a spatial inversion.

\subsection{Generalization to arbitrary angles}
\label{sec:angles}
As mentioned, it is straightforward to generalize the derivation of
the tight-binding equations to arbitrary bonding angles
$\alpha\neq \pi/4$ in the
zigzag chain. We just outline the main steps: In (\ref{expansion}) and
the following, we replace $y'-(-1)^n d'/2$ with $y'-(-1)^n  \tan (\alpha) d'/2$
(having redefined $d'=d \cos \alpha$). In (\ref{trig}), (\ref{sigmaprime}),
$\pi/4$ then get replaced by $\alpha$, as already indicated.
In (\ref{diag}) we get $e^{-i 2 \alpha}\sigma$ for diagonal links and
$e^{i 2 \alpha}\sigma$ for
antidiagonal, and in (\ref{diag2}) we get $e^{i 2 \alpha}\sigma$
for diagonal links and
$e^{-i 2 \alpha}\sigma$ for
antidiagonal. Then in the tight-binding equations of motion
(\ref{uvdnls}), the last term for the spin-up component gets replaced by
$+e^{(-1)^n i 2 \alpha} v_{n+1} - e^{-(-1)^n i 2 \alpha} v_{n-1}$, and for the spin-down
component by $+e^{-(-1)^n i 2 \alpha} u_{n+1} - e^{(-1)^n i 2 \alpha} u_{n-1}$.
For the rest of this paper we will assume $\alpha = \pi/4$ and leave the study
of effects of variation of the binding angle to future work.

\subsection{Linear dispersion relation}
\label{sec:disp}

\begin{figure}
\includegraphics[height=0.33\textwidth,angle=270]{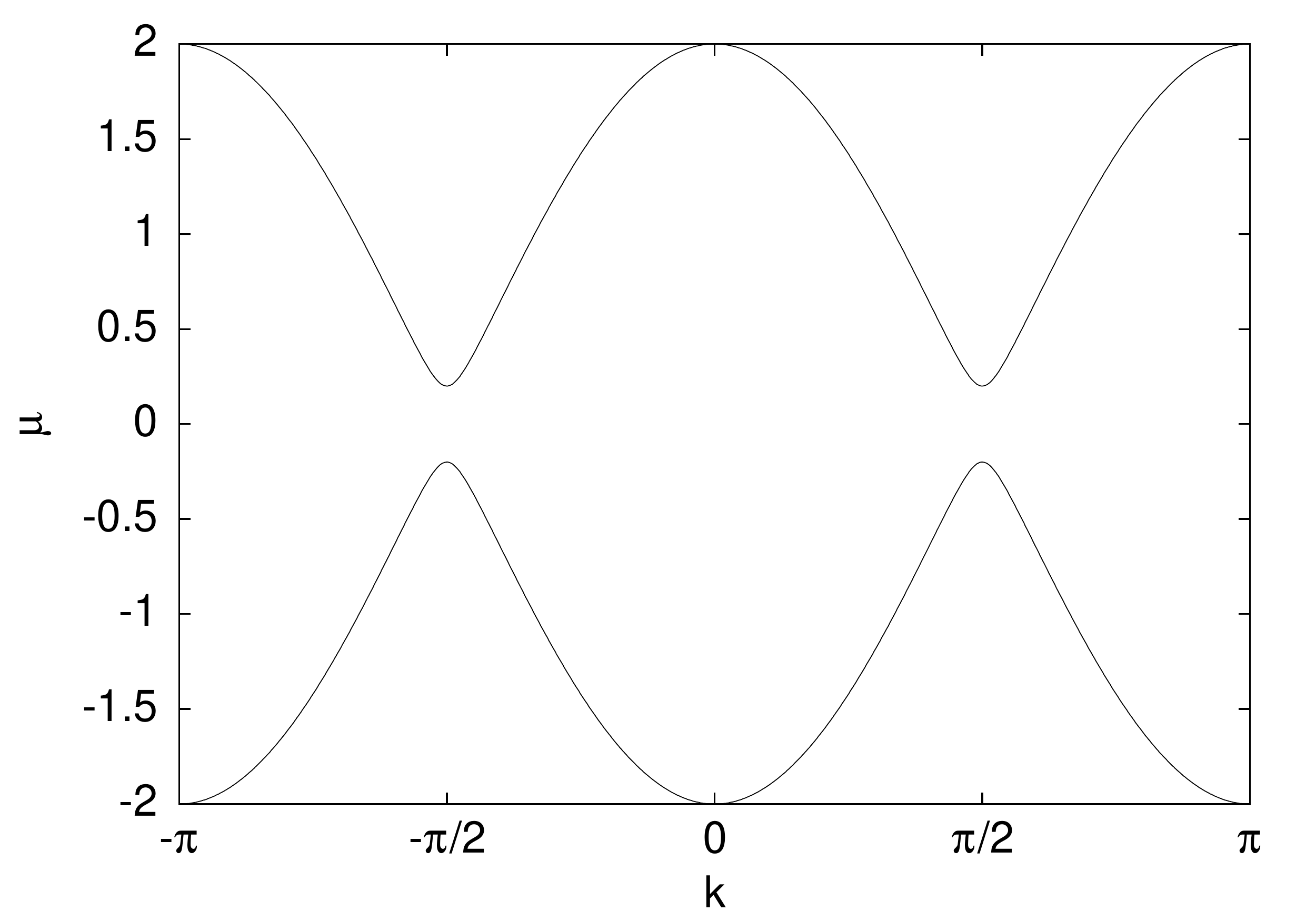}
\caption{The dispersion relation (\ref{disp}) with
  $\Gamma=1$ and $\sigma=0.1$.
 }
\label{fig:disp}
\end{figure}

Let $u_n = u e^{i(kn-\mu t)}, v_n=v e^{i[(k+\pi)n-\mu t]}$ (i.e.,
$v_n'= v e^{i(kn-\mu t)}$ removing factors $(-1)^n$), with
$|u|, |v| \ll 1$. Inserting it into (\ref{uvdnls}) (or (\ref{uvprime})) and
neglecting the nonlinear terms then yields:
\begin{equation}
\mu_{1,2}(k) = \pm 2 \sqrt{\Gamma^2 \cos^2 k + \sigma^2 \sin^2k}.
\label{disp}
\end{equation}
Thus, as illustrated in Fig.\ \ref{fig:disp} (assuming $\sigma < \Gamma$),
the spin-orbit coupling opens up gaps
in the linear dispersion relation at $k=\pm \pi/2$, of width $4\sigma$. Note
that in contrast to the models for straight chains studied in
\cite{Belicev15,Salerno15}, no external magnetic field is needed to open
the gap for the zigzag chain. The gap opening is a consequence of the
simultaneous presence of spin-orbit coupling and nontrivial geometry, which
was also noted for the more complicated diamond chain in \cite{Gligoric16}.

The amplitude ratios between the components
may be obtained as
$v/u=\frac{-\Gamma \cos k \mp \sqrt{\Gamma^2 \cos^2 k + \sigma^2 \sin^2k}}
{\sigma \sin k}$.
For weak spin-orbit coupling  ($\sigma \ll \Gamma$), the polariton is mainly
spin-up ($u\gg v$) on the lower dispersion branch and spin-down on the upper
branch
($v\gg u$) when $\Gamma \cos k >0$, and the opposite when $\Gamma \cos k <0$.

\subsection{Flat band and compact modes}
\label{sec:flat}
We may also note that in the particular case of
$|\Gamma| = |\sigma|$, the dispersion relation becomes exactly flat.
In this case, there are eigenmodes completely localized on either upper
or lower part of the chain, with alternating $v_n=\pm i u_n$ on this part
(i.e., $v_n\equiv u_n\equiv 0$ either for odd or even $n$).
These modes persist also in the presence of nonlinearity (interactions).
With the flat band, it is also possible to construct exact compact solutions
localized on two neighboring sites. Explicitly, we get for $\sigma = + \Gamma$:
\begin{equation}
\left(\begin{array}{c}
  u_{n_0}\\v_{n_0}'\end{array}
  \right) =
  A e^{-i\mu t}
\left(\begin{array}{c}
  1\\i\end{array}\right);
  \left(\begin{array}{c}
  u_{n_0+1}\\v_{n_0+1}'\end{array}
\right) =\pm A e^{-i\mu t}
\left(\begin{array}{c}
  1\\-i  \end{array}\right) ,
\label{eq:compacton+}
\end{equation}
and  for $\sigma = - \Gamma$:
\begin{equation}
\left(\begin{array}{c}
  u_{n_0}\\v_{n_0}'\end{array}
  \right) =
  A e^{-i\mu t}
\left(\begin{array}{c}
  1\\-i\end{array}\right);
  \left(\begin{array}{c}
  u_{n_0+1}\\v_{n_0+1}'\end{array}
\right) =\pm A e^{-i\mu t}
\left(\begin{array}{c}
  1\\i  \end{array}\right) .
\label{eq:compacton-}
\end{equation}
In both cases, the nonlinear dispersion relation for these compactons yields
$\mu = (1+\mathfrak{a}) |A|^2 \mp 2 \Gamma$.
We will discuss further properties of these nonlinear compactons
  (e.g. stability) below.

Before proceeding, we briefly discuss some connections between our results 
above and earlier studies of compact flat-band modes in different contexts. 
(See, e.g., Ref.\ \cite{Derzhko} for an extensive review of earlier results 
on flat-band modes in spin systems and strongly correlated electron models, 
and Refs.\ \cite{Leykam1, Leykam2} for reviews of more recent experimental 
and theoretical progress.) As regards the properties in the linear 
flat-band limit, 
our model belongs to the same class of models as those describing hopping 
between $s$- and $p$-orbital states, e.g., the ``topological orbital ladders'' 
proposed in Ref.\ \cite{Li} for ultracold atoms in higher orbital bands. 
In the general classification scheme of compact localized flat-band modes 
occupying two unit cells in a one-dimensional nearest-neighbour coupled 
lattice, the relevant case is that described 
in Appendix B3 of Ref.\ \cite{Maimaiti} with two coexisting, 
nondegenerate, flat bands. 
As far as we are aware, the corresponding {\em nonlinear} compact modes have 
not been investigated in any earlier work. By contrast, there are several 
works studying nonlinear compactons in a 'sawtooth' lattice 
\cite{Naether,Danieli} which would 
result if an additional next-nearest neighbour (horizontal) coupling was added 
to {\em either} the upper {\em or} the lower sub-chain (but not both) in 
Fig.\ \ref{fig:zig-zag}. In this case, compactons may appear without presence 
of spin-orbit coupling, instead due to balance between nearerst and 
next-nearest 
neighbor couplings. For the sawtooth chain, one of the two bands will always 
remains dispersive.

\section{Nonlinear localized modes}
\label{sec:nonlinear}
\subsection{Single-site modes above the spectrum in the weak-coupling limit}

For the case of no spin-orbit coupling ($\sigma=0$ and small $\Gamma \ll 1$),
analysis of fundamental nonlinear localized solutions (including their
linear stability) of (\ref{uvdnls}) was done in \cite{Darmanyan98}. It would
be straightforward to redo a similar extensive analysis
including also a small $\sigma$, but it is not the main aim of this work.
We focus here first on discussing the effect of small
coupling on polaritons with main localization on a single site $n_0$.

In the limit of $\Gamma=\sigma=0$ (``anticontinuous limit''), stationary
solutions
of (\ref{uvprime}) are
well known. There are two spin-polarized solutions:
$\left(\begin{array}{c}
  u_{n_0}\\v_{n_0}\end{array}
\right) = \sqrt{\mu}e^{-i\mu t}
\left(\begin{array}{c}
  1\\0\end{array}\right)
  $
  (spin-up);
  $\left(\begin{array}{c}
  u_{n_0}\\v_{n_0}\end{array}
\right) = \sqrt{\mu}e^{-i\mu t}
\left(\begin{array}{c}
  0\\1\end{array}\right)
  $
  (spin-down); and one spin-mixed solution:
$\left(\begin{array}{c}
  u_{n_0}\\v_{n_0}\end{array}
\right) = \sqrt{\frac{\mu}{1+\mathfrak{a}}}e^{-i\mu t}
\left(\begin{array}{c}
  1\\e^{i\theta}\end{array}\right)
$,
with an arbitrary relative phase $\theta$ between the spin components.
Comparing the Hamiltonian (\ref{Ham}) for these solutions at given norm
$P$, we have $H=P^2/2$ for the spin-polarized modes and $H=(1+\mathfrak{a})P^2/4$
for the mixed mode,
so the mixed mode has lowest energy as long as $\mathfrak{a}<1$.

When
$\mu>0$ does not belong to the linear spectrum (\ref{disp}), we search
for continuation of these modes for small but nonzero $\Gamma, \sigma$
into nonlinear
localized modes with exponentially decaying tails and frequency above the
spectrum.
(We here assume $\mathfrak{a} > -1$; if $\mathfrak{a} < -1$ the localized modes
arising from the spin-mixed solution will have $\mu <0$ and thus lie below
the spectrum.)
We may
calculate them explicitly perturbatively to arbitrary order in the two small
parameters $\Gamma, \sigma$; here we give only the first- and second-order
corrections to the five central sites (amplitudes of other sites will be
of higher order):
\begin{eqnarray}
\left(\begin{array}{c}
  u_{n_0}\\v_{n_0}'\end{array}
  \right) \approx
  \left(\sqrt{\mu}-\frac{\Gamma^2+\sigma^2}{\mu^{3/2}}\right)e^{-i\mu t}
\left(\begin{array}{c}
  1\\0\end{array}\right);
  \left(\begin{array}{c}
  u_{n_0\pm1}\\v_{n_0\pm1}'\end{array}
\right) \approx \frac{1}{\sqrt{\mu}}e^{-i\mu t}
\left(\begin{array}{c}
  -\Gamma\\\pm i \sigma \end{array}\right); \nonumber \\
  \left(\begin{array}{c}
  u_{n_0\pm2}\\v_{n_0\pm2}'\end{array}
\right) \approx \frac{\Gamma^2-\sigma^2}{\mu^{3/2}}e^{-i\mu t}
\left(\begin{array}{c}
  1\\0\end{array}\right) \quad \text{(``spin-up'')};
\label{eq:spin-up}
\end{eqnarray}
\begin{eqnarray}
\left(\begin{array}{c}
  u_{n_0}\\v_{n_0}'\end{array}
  \right) \approx
  \left(\sqrt{\mu}-\frac{\Gamma^2+\sigma^2}{\mu^{3/2}}\right)e^{-i\mu t}
\left(\begin{array}{c}
  0\\1\end{array}\right);
  \left(\begin{array}{c}
  u_{n_0\pm1}\\v_{n_0\pm1}'\end{array}
\right) \approx \frac{1}{\sqrt{\mu}}e^{-i\mu t}
\left(\begin{array}{c}
  \pm i \sigma \\\Gamma \end{array}\right); \nonumber \\
  \left(\begin{array}{c}
  u_{n_0\pm2}\\v_{n_0\pm2}'\end{array}
\right) \approx \frac{\Gamma^2-\sigma^2}{\mu^{3/2}}e^{-i\mu t}
\left(\begin{array}{c}
  0\\1\end{array}\right) \quad \text{(``spin-down'')};
\label{eq:spin-down}
\end{eqnarray}
\begin{eqnarray}
\left(\begin{array}{c}
  u_{n_0}\\v_{n_0}'\end{array}
  \right) \approx
  \frac{\mu^2-(\Gamma^2+\sigma^2)}{\sqrt{\mu^{3}(1+\mathfrak{a})}}e^{-i\mu t}
\left(\begin{array}{c}
  1\\e^{i \theta}\end{array}\right);
  \left(\begin{array}{c}
  u_{n_0\pm1}\\v_{n_0\pm1}'\end{array}
\right) \approx \frac{1}{\sqrt{\mu(1+\mathfrak{a})}}e^{-i\mu t}
\left(\begin{array}{c}
  -\Gamma \pm i \sigma e^{i \theta}\\\Gamma e^{i \theta} \pm i \sigma
\end{array}\right);\nonumber \\
  \left(\begin{array}{c}
  u_{n_0\pm2}\\v_{n_0\pm2}'\end{array}
\right) \approx \frac{\Gamma^2-\sigma^2}{\sqrt{\mu^{3}(1+\mathfrak{a})}}e^{-i\mu t}
\left(\begin{array}{c}
  1\\e^{i \theta}\end{array}\right) \quad \text{(``spin-mixed'')}.
\label{eq:mixed}
\end{eqnarray}
It can be seen from such expressions (extending to higher orders)
that amplitudes do decay
exponentially above the spectrum, $\mu>2 \Gamma$. However, for
spin-mixed modes with $|u_n| = |v_n'|$, it is important to remark
that even
though the second-order corrections in (\ref{eq:mixed}) can be obtained for
arbitrary relative phases $\theta$, the fourth-order correction to site
$n_0$ can be made consistent with the condition $|u_n| = |v_n'|$
only if $\Gamma^2 \sigma^2 \sin (2 \theta) = 0$. Thus, since a solution
with $\theta = \pi$ is equivalent to $\theta=0$ through spatial reflection
in the central site, the only non-equivalent single-site
centered spin-mixed  modes existing
for nonzero $\Gamma$ and $\sigma$ have $\theta=0, \pi/2$.
We also remark that, for a stationary and localized solution, current
conservation imposes the general condition:
\begin{equation}
\Gamma \Im (u_{n+1}^*u_n - v_{n+1}'^*v_n') =
\sigma \Re (v_{n+1}'^*u_n + u_{n+1}^*v_n') .
\label{eq:current}
\end{equation}
\begin{figure}
\includegraphics[width=12cm]{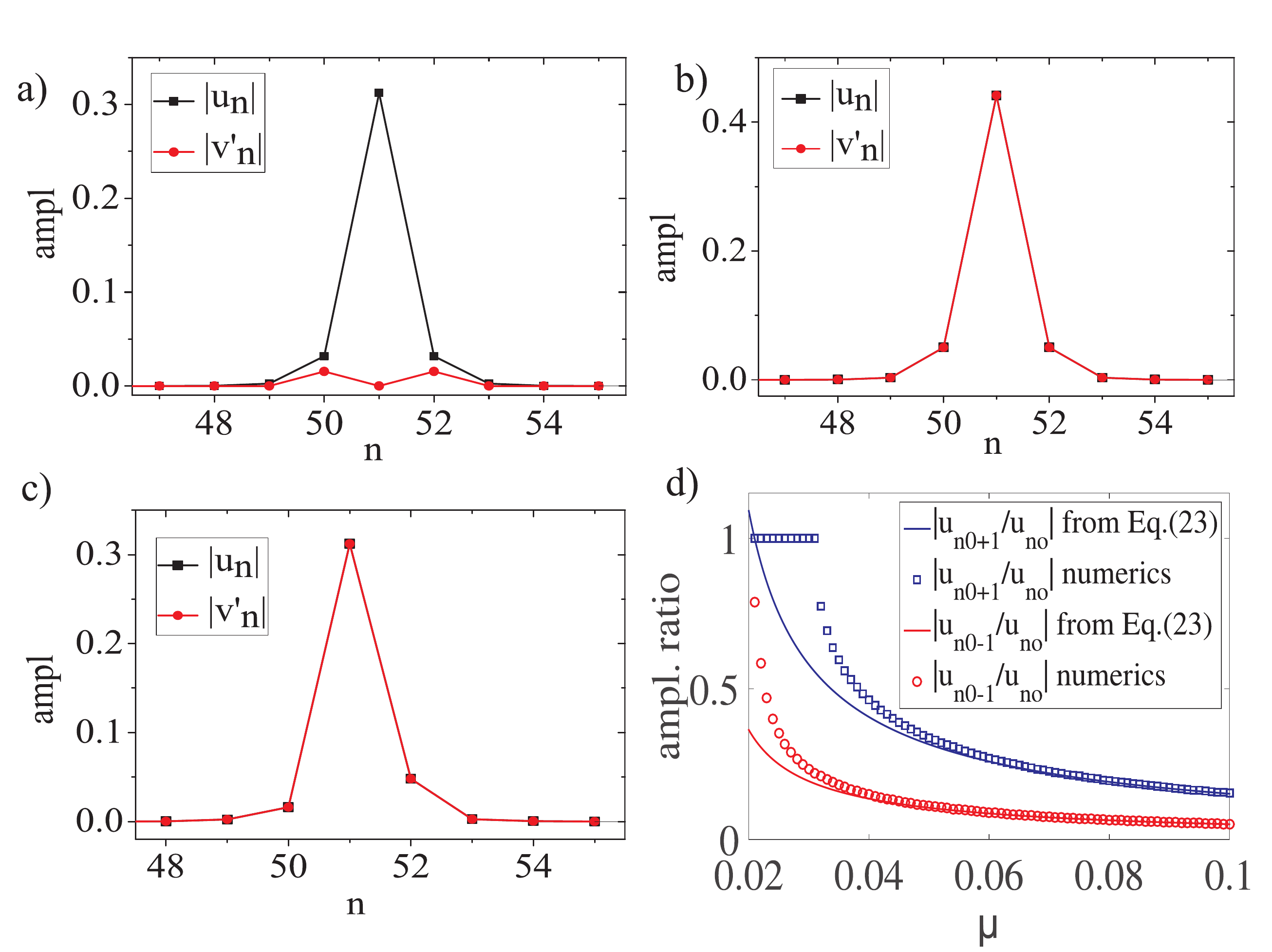}
\caption{
Numerical examples of fundamental nonlinear localized modes in the
semi-infinite gap above the linear spectrum when
$\Gamma = 0.01$, $\sigma=0.005$, and $\mu=0.1$:
 spin-up mode (\ref{eq:spin-up}) (a);
spin-mixed mode (\ref{eq:mixed}) when
$\mathfrak{a} = -0.5$ and $\theta=0$ (b);
and spin-mixed mode (\ref{eq:mixed}) when $\mathfrak{a} = 0 $ and
$\theta=\pi/2$ (c). Amplitude ratios between central and two
neighboring
sites obtained from numerics and Eq.\ (\ref{eq:mixed}) with $\theta = \pi/2$
for the continuation of the solution in (c) 
towards smaller $\mu$ are shown in (d).}
\label{fig:up_mixed}
\end{figure}

Numerically calculated
examples for the spin-up and spin-mixed modes are illustrated in
Fig.\ \ref{fig:up_mixed}.
Note from (\ref{eq:mixed}) that, for the spin-mixed mode with $\theta = \pi/2$,
$|u_{n_0+1}|^2 +|v_{n_0+1}'|^2 \neq|u_{n_0-1}|^2 +|v_{n_0-1}'|^2  $, i.e., the
reflection symmetry around the central site gets broken on the opposite
sublattice (upper or lower part of the chain) if there is a nontrivial
phase-shift between the spin-up and spin-down components at the central site.
As $|\Gamma \sigma| /\mu^2$ increases the spatial asymmetry increases
(Fig.\ \ref{fig:up_mixed} (d)), until the solution typically
bifurcates with an
inter-site centered (two-site) mode with equal amplitudes at sites
$n_0$ and $n_0+1$ before reaching the upper band edge at $\mu = 2 \Gamma$.

\subsection{Fundamental gap modes from the flat-band limit}

\begin{figure}
\includegraphics[width=12cm]{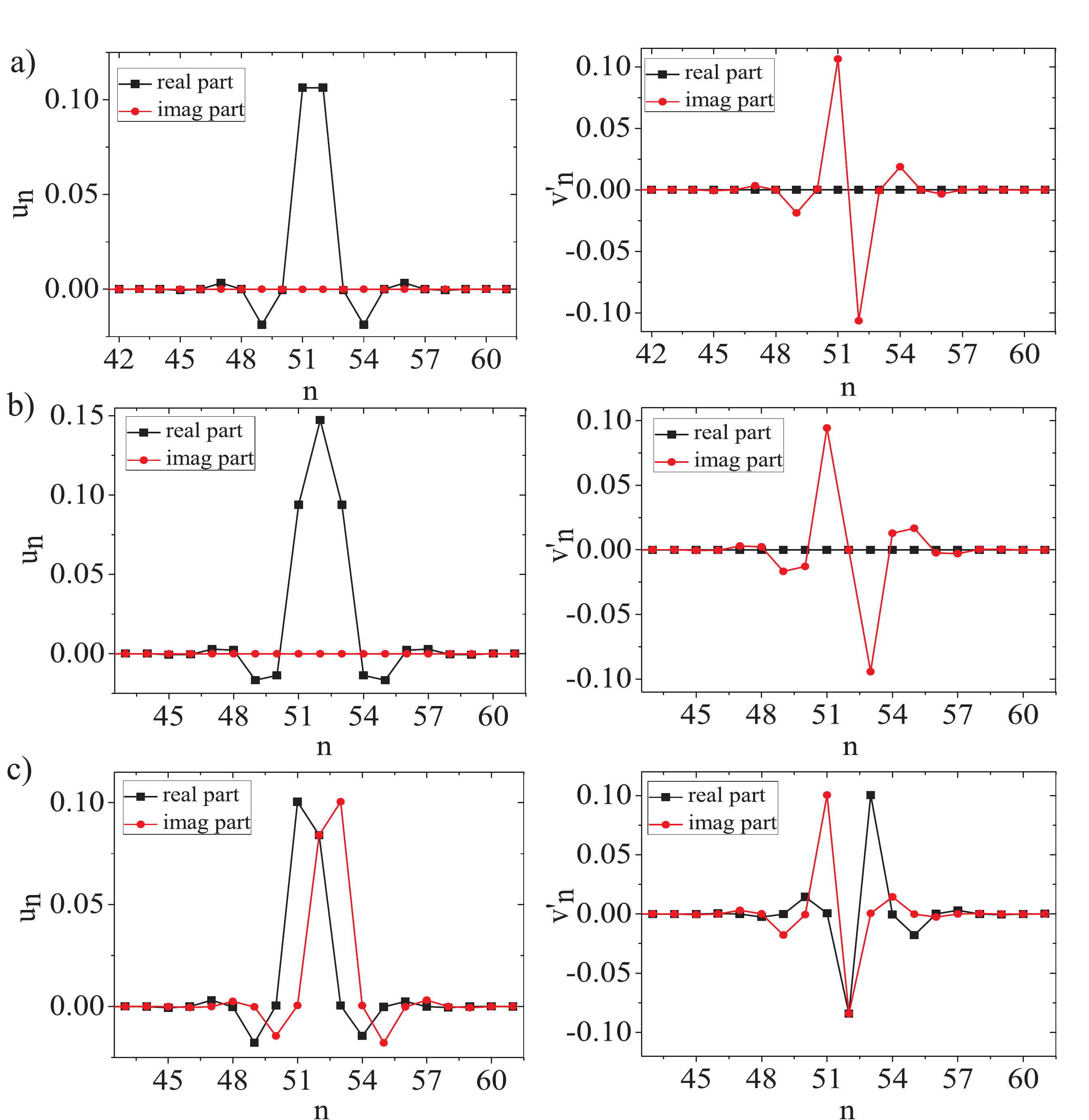}
\caption{Numerical examples of $u_{n}$ and $v^{,}_{n}$ components of fundamental (type I) (a), type II (b) and type III (c) gap modes found in the
mini-gap opened by the spin-orbit coupling when
$\Gamma = 0.01$, $\sigma=0.007$, $\mathfrak{a}=0.5$, and $\mu=0$.}
\label{fig:gap_mode}
\end{figure}
Since the gap in the linear spectrum opened by the
spin-orbit coupling at $k = \pm \pi/2$ appears only when $\Gamma$ and
$\sigma$ are both nonzero, the
standard anticontinuous limit  $\Gamma = \sigma = 0$ is not suitable for
constructing nonlinear localized modes with frequency inside this gap
(``discrete gap solitons''). Instead, we may use the flat-band limit
$|\Gamma| = |\sigma| \neq 0$, where the exact nonlinear compacton modes
(\ref{eq:compacton+})-(\ref{eq:compacton-}) can be used as ``building blocks''
for the continuation procedure. Analogously to above, we may then calculate
gap solitons perturbatively in the small
parameter $|\Gamma|-|\sigma|$. To be specific, we assume $\mathfrak{a} > -1$,
$\Gamma \geq \sigma > 0$, and consider the continuation of a single two-site
compacton from the lower flat band $\mu=-2\Gamma$ into the gap. From the
limiting solution (\ref{eq:compacton+}) with the upper sign, we then obtain
the lowest-order corrections to six central sites (amplitudes at other sites
are of higher order) as:
\begin{eqnarray}
\left(\begin{array}{c}
  u_{n_0}\\v_{n_0}'\end{array}
  \right) \approx
  \sqrt{\frac{\mu+2\Gamma}{1+\mathfrak{a}}}
\left(1-\frac{\Gamma - \sigma}{5 \Gamma - \sigma +2 \mu}\right)e^{-i\mu t}
\left(\begin{array}{c}
  1\\{i }\end{array}\right);
\left(\begin{array}{c}
  u_{n_0+1}\\v_{n_0+1}'\end{array}
  \right) \approx
  \sqrt{\frac{\mu+2\Gamma}{1+\mathfrak{a}}}
\left(1-\frac{\Gamma - \sigma}{5 \Gamma - \sigma +2 \mu}\right)e^{-i\mu t}
\left(\begin{array}{c}
  1\\{-i }\end{array}\right); \nonumber \\
  \left(\begin{array}{c}
  u_{n_0-1}\\v_{n_0-1}'\end{array}
\right) \approx
\sqrt{\frac{\mu+2\Gamma}{1+\mathfrak{a}}}
\frac{\mu (\Gamma - \sigma)}{(\Gamma + \sigma)^2 -\mu^2}e^{-i\mu t}
\left(\begin{array}{c}
  1\\{-i }\end{array}\right);
  \left(\begin{array}{c}
  u_{n_0+2}\\v_{n_0+2}'\end{array}
\right) \approx
  \sqrt{\frac{\mu+2\Gamma}{1+\mathfrak{a}}}
\frac{\mu (\Gamma - \sigma)}{(\Gamma + \sigma)^2 -\mu^2}e^{-i\mu t}
\left(\begin{array}{c}
  1\\{i }\end{array}\right); \nonumber \\
  \left(\begin{array}{c}
  u_{n_0-2}\\v_{n_0-2}'\end{array}
\right) \approx
- \sqrt{\frac{\mu+2\Gamma}{1+\mathfrak{a}}}
\frac{\Gamma^2 - \sigma^2}{(\Gamma + \sigma)^2-\mu^2}e^{-i\mu t}
\left(\begin{array}{c}
  1\\{i }\end{array}\right)
  \left(\begin{array}{c}
  u_{n_0+3}\\v_{n_0+3}'\end{array}
\right) \approx
  - \sqrt{\frac{\mu+2\Gamma}{1+\mathfrak{a}}}
\frac{\Gamma^2 - \sigma^2}{(\Gamma + \sigma)^2-\mu^2}e^{-i\mu t}
\left(\begin{array}{c}
  1\\{-i }\end{array}\right) .
\label{eq:gap}
\end{eqnarray}
This family of fundamental gap modes (called type I gap modes) can be continued throughout the gap,
with a numerical example illustrated in Fig.\ \ref{fig:gap_mode} (a). Profiles of another two types of gap modes 
numerically found to exist as nonlinear continuation of fundamental compactons, are depicted in Fig.\ \ref{fig:gap_mode} (b,c). 
Family of gap modes of type II (Fig.\ \ref{fig:gap_mode} (b)) originates from compact solution which is superposition 
of two neighboring overlapping in-phase compactons. On the other hand, type III gap modes evolve in the presence of 
nonlinearity from superposition of two neighboring overlapping compactons 
with a $\pi/2$ phase difference (Fig.\ \ref{fig:gap_mode} (c)).  

\section{Linear stability of nonlinear localized modes}
\label{sec:stab}
Linear stability of the above modes can be checked from the standard
eigenvalue problem. If we denote the amplitudes of the exact stationary modes of
(\ref{uvprime}) as
$\{u_n^{(0)}, v_n'^{(0)}\}$, we may express the perturbed modes as
$u_n = \left[u_n^{(0)}+(c_ne^{-i\lambda t} + d_n^* e^{i \lambda^*t})\right]e^{-i\mu t} $,
$v_n' = \left[v_n'^{(0)}+(f_ne^{-i\lambda t} + g_n^* e^{i \lambda^*t})\right]e^{-i\mu t} $.
Inserting into (\ref{uvprime}) and linearizing, we obtain the following
linear system of equations for the perturbation amplitudes 
$\{c_n, d_n, f_n, g_n\}$:
\begin{eqnarray}
  \left(-\mu+2|u_n^{(0)}|^2+\mathfrak{a} |v_n'^{(0)}|^2\right)c_n + u_n^{(0)2} d_n +
  \mathfrak{a} u_n^{(0)}v_n'^{(0)*} f_n + \mathfrak{a} u_n^{(0)} v_n'^{(0)} g_n
  -\Gamma (c_{n+1}+c_{n-1})-i\sigma(f_{n+1}-f_{n-1})=\lambda c_n
  \nonumber \\
  \left(\mu-2|u_n^{(0)}|^2-\mathfrak{a} |v_n'^{(0)}|^2\right)d_n - u_n^{(0)*2} c_n -
  \mathfrak{a} u_n^{(0)*}v_n'^{(0)*} f_n - \mathfrak{a} u_n^{(0)*} v_n'^{(0)} g_n
  +\Gamma (d_{n+1}+d_{n-1})-i\sigma(g_{n+1}-g_{n-1})=\lambda d_n
  \nonumber \\
  \left(-\mu+2|v_n'^{(0)}|^2+\mathfrak{a} |u_n^{(0)}|^2\right)f_n + v_n'^{(0)2} g_n +
  \mathfrak{a} u_n^{(0)*}v_n'^{(0)} c_n + \mathfrak{a} u_n^{(0)} v_n'^{(0)} d_n
  +\Gamma (f_{n+1}+f_{n-1})-i\sigma(c_{n+1}-c_{n-1})=\lambda f_n
 \nonumber \\
  \left(\mu-2|v_n'^{(0)}|^2-\mathfrak{a} |u_n^{(0)}|^2\right)g_n - v_n'^{(0)*2} f_n -
  \mathfrak{a} u_n^{(0)*}v_n'^{(0)*} c_n - \mathfrak{a} u_n^{(0)} v_n'^{(0)*} d_n
  -\Gamma (g_{n+1}+g_{n-1})-i\sigma(d_{n+1}-d_{n-1})=\lambda g_n .
  \nonumber\\
  \label{stability}
\end{eqnarray}
Linear stability is then equivalent to (\ref{stability}) having no
complex eigenvalues. We may easily solve it for the uncoupled modes.
Due to the overall gauge invariance of (\ref{uvprime})
($u_n\rightarrow e^{i\phi}u_n,v_n'\rightarrow e^{i\phi}v_n'$), there are always
two eigenvalues at $\lambda=0$. For the spin-polarized modes, the remaining
two eigenvalues are at $\lambda=\pm(1-\mathfrak{a})\mu$, while for the spin-mixed mode
there is a fourfold degeneracy at $\lambda=0$. The latter is explained by
the arbitrary phase difference $\theta$ between the $u$ and $v$ components
for this mode.

\begin{figure}
\includegraphics[width=0.33\textwidth]{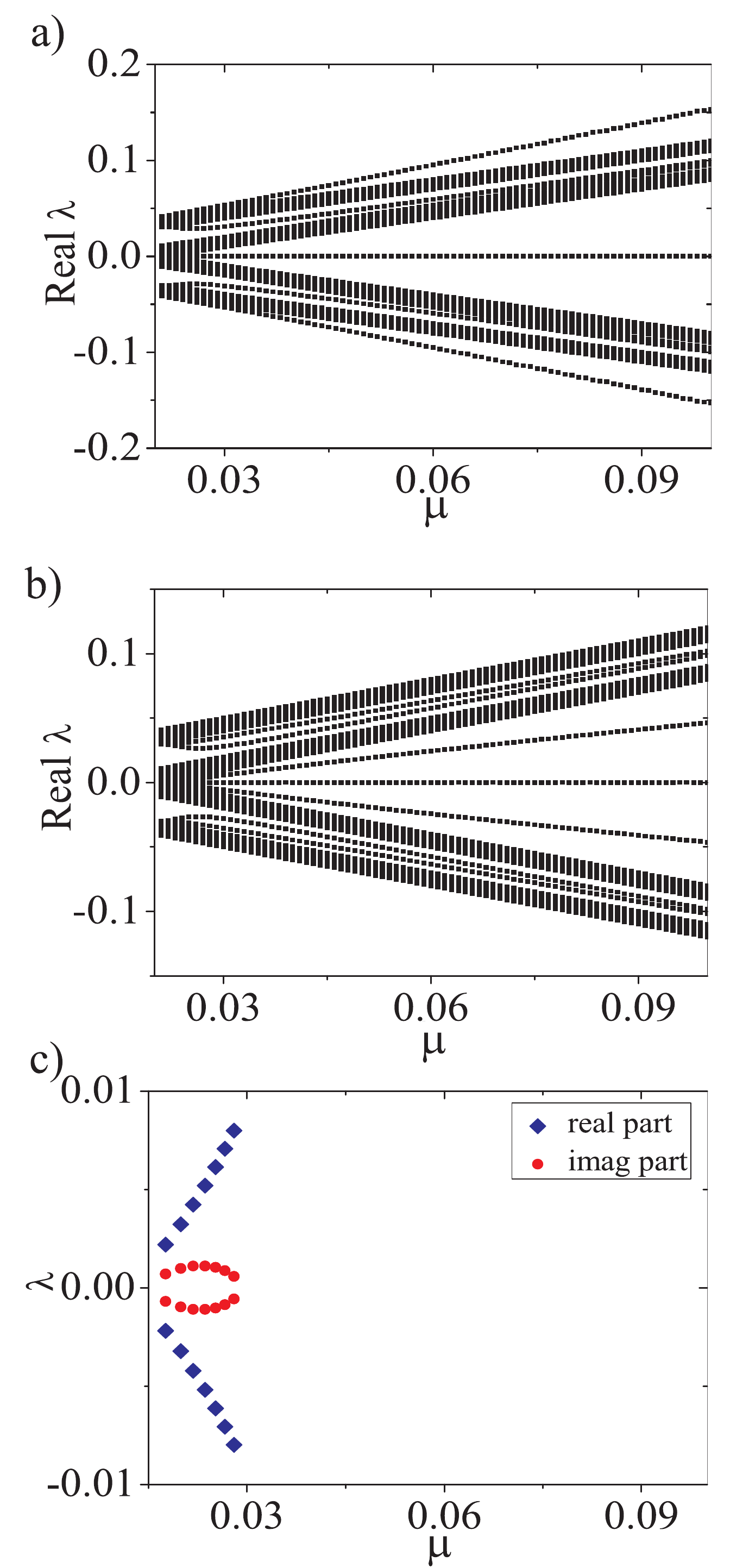}
\caption{Stability eigenvalues for the continuation of
fundamental spin-up states (\ref{eq:spin-up}) when
$\Gamma = 0.01$, $\sigma=0.005$, and $\mathfrak{a}=-0.5$ (a),
$\mathfrak{a}=0.5$ (b), and $\mathfrak{a}=1.5$ (c), respectively.
For $\mathfrak{a} \leq 1$ the imaginary parts of eigenvalues are
zero to numerical accuracy. Only the unstable eigenvalues are shown for
$\mathfrak{a}=1.5$. }
\label{fig:stab_up}
\end{figure}
To see whether linear stability of the fundamental modes survives switching on
the couplings
$\Gamma, \sigma$, we first note that the linear spectrum of (\ref{stability})
corresponding to sites with $u_n^{(0)}\equiv v_n'^{(0)}\equiv 0$ has four
branches, at $\lambda \in \pm[\mu-2\Gamma, \mu - 2\sigma]$ and
$\lambda \in \pm[\mu+2\sigma, \mu + 2\Gamma]$. Thus, unless
$\mathfrak{a}=0, 1,\, \text{or}\, 2$, we see immediately that  the
fundamental spin-polarized
modes must remain linearly stable at least for small couplings.
The general stability properties
for larger $\Gamma$ and/or $\sigma$ will be discussed below for the
different fundamental modes separately.

\subsection{Spin-polarized modes above the spectrum}

Typical results  from numerical diagonalization of (\ref{stability}) for
the family of fundamental spin-polarized modes above the spectrum
are shown in Fig.\ \ref{fig:stab_up}.
As is seen, these modes are
{\em linearly stable in their full regime of existence}
when $\mathfrak{a} < 1$.
The magnitude of the frequency of the internal eigenmode arising from local
oscillations at the central site lies  above the linear
spectrum when $\mathfrak{a} < 0$ (Fig.\ \ref{fig:stab_up} (a)) and below the
linear spectrum when  $0 < \mathfrak{a} < 1$ (Fig.\ \ref{fig:stab_up} (b)).
In
both cases, it smoothly joins the band edge as $\mu \rightarrow 2 \Gamma$
(linear limit), without causing any resonances.
On the other hand, for $\mathfrak{a} >1$, the Krein signature of this
eigenmode will change, as a consequence of the spin-polarized mode now having
a lower energy than a spin-mixed mode, and thus it is no longer an energy
maximizer for the system. This results in small regimes of weak
oscillatory instabilities
when the internal mode collides with the linear spectrum for frequencies close
to the band edge, as shown in Fig.\ \ref{fig:stab_up} (c).

\subsection{Spin-mixed modes above or below the spectrum}
\begin{figure}
\includegraphics[width=12cm]{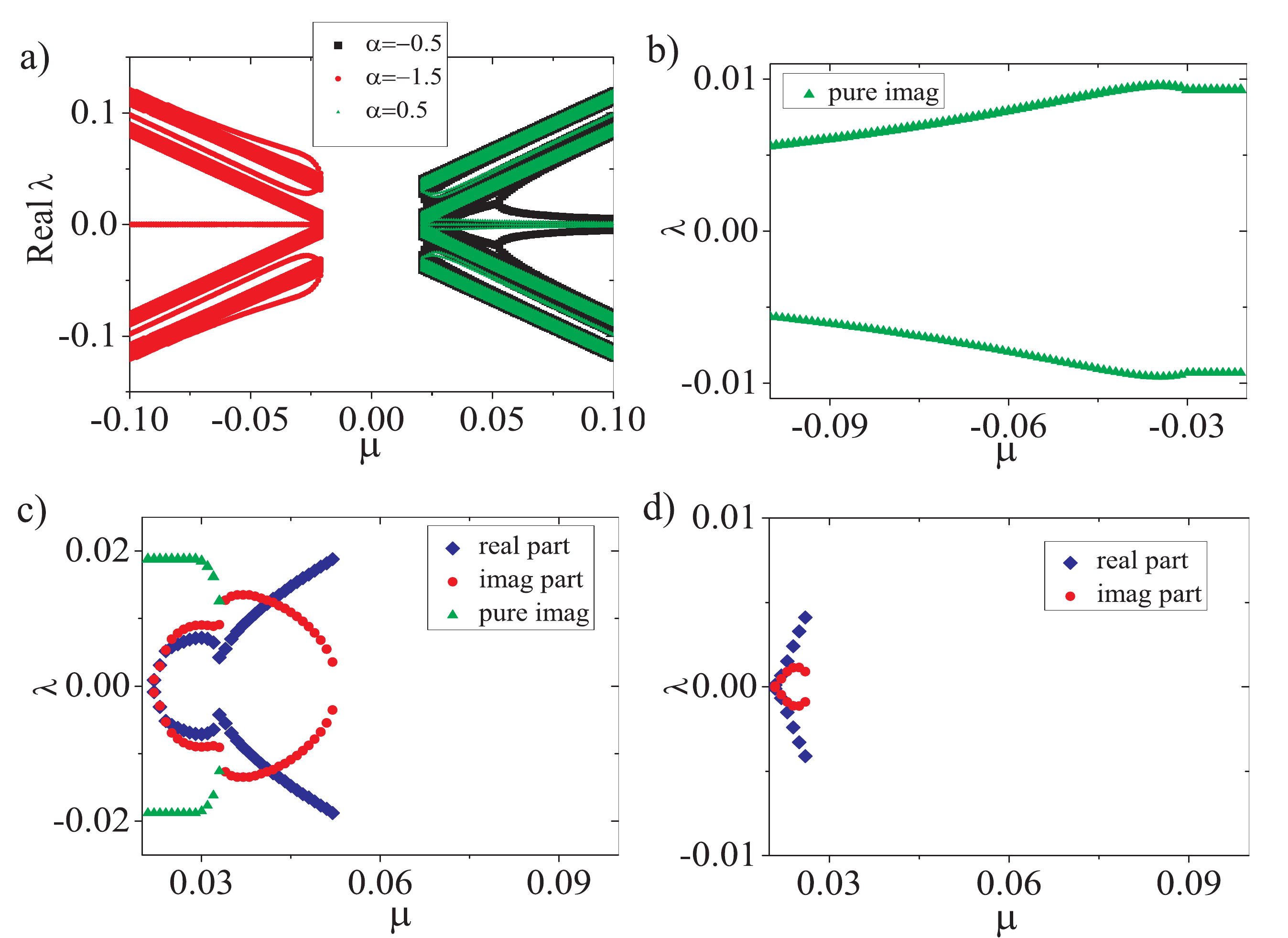}\\
\caption{Stability eigenvalues for the continuation of
fundamental spin-mixed states (\ref{eq:mixed}) with $\theta = 0$ when
$\Gamma = 0.01$ and $\sigma=0.005$. (a) Real parts of
eigenvalues when $\mathfrak{a}=-1.5$ (red (middle gray) circles),
$\mathfrak{a}=-0.5$ (black squares),
and $\mathfrak{a}=0.5$ (green (light gray) triangles), respectively. Unstable eigenvalues 
when  $\mathfrak{a}=-1.5$ (b),
$\mathfrak{a}=-0.5$ (c),
and $\mathfrak{a}=0.5$ (d), respectively. Here, purely imaginary
eigenvalues are represented by green (light gray) triangles, and complex eigenvalues are
represented by blue (dark gray) squares and red (middle gray) circles for their real and imaginary
parts, respectively.}
\label{fig:stab_mixed}
\end{figure}

For the fundamental spin-mixed modes continued from (\ref{eq:mixed}), the
four-fold degeneracy of zero eigenvalues resulting from the relative phase
$\theta$ is generally
broken for non-zero coupling as only modes with integer $2\theta/\pi$ can
be continued, and moreover the structures of modes with $\theta = 0$ and
$\theta = \pi/2$ become
non-equivalent. We discuss here first the case $\theta = 0$, and
show in Fig.\ \ref{fig:stab_mixed} typical results  from numerical
diagonalization for different values of $\mathfrak{a}$.
First, for $\mathfrak{a} < -1$, as remarked above the spin-mixed modes lie
below the linear spectrum ($\mu < -2 \Gamma $), and the pair of eigenvalues
originating
from $\lambda = 0$ in the anticontinuous limit ($\mu \rightarrow - \infty$)
generally
goes out along the imaginary axis (Fig.\ \ref{fig:stab_mixed} (b)),
where it remains. Thus, spin-mixed modes with
$\theta=0$ and $\mathfrak{a} < -1$ are generically unstable.
On the other hand, when  $\mathfrak{a} > -1$ the spin-mixed modes  lie
above the linear spectrum ($\mu > 2 \Gamma $), and for $-1 < \mathfrak{a} < 1$
 this eigenvalue pair goes
out along the real axis (Fig.\ \ref{fig:stab_mixed} (a)). Thus, these
modes remain linearly stable for sufficiently large $\mu$ (or, equivalently,
weak coupling), but become unstable through oscillatory instabilities
(complex eigenvalues, see Figs.\ \ref{fig:stab_mixed} (c,d)) as they approach the linear band edge with widening
tails, causing resonances between the local oscillation mode at the central
site and modes arising from oscillations at small-amplitude sites.

\begin{figure}
\includegraphics[width=8cm]{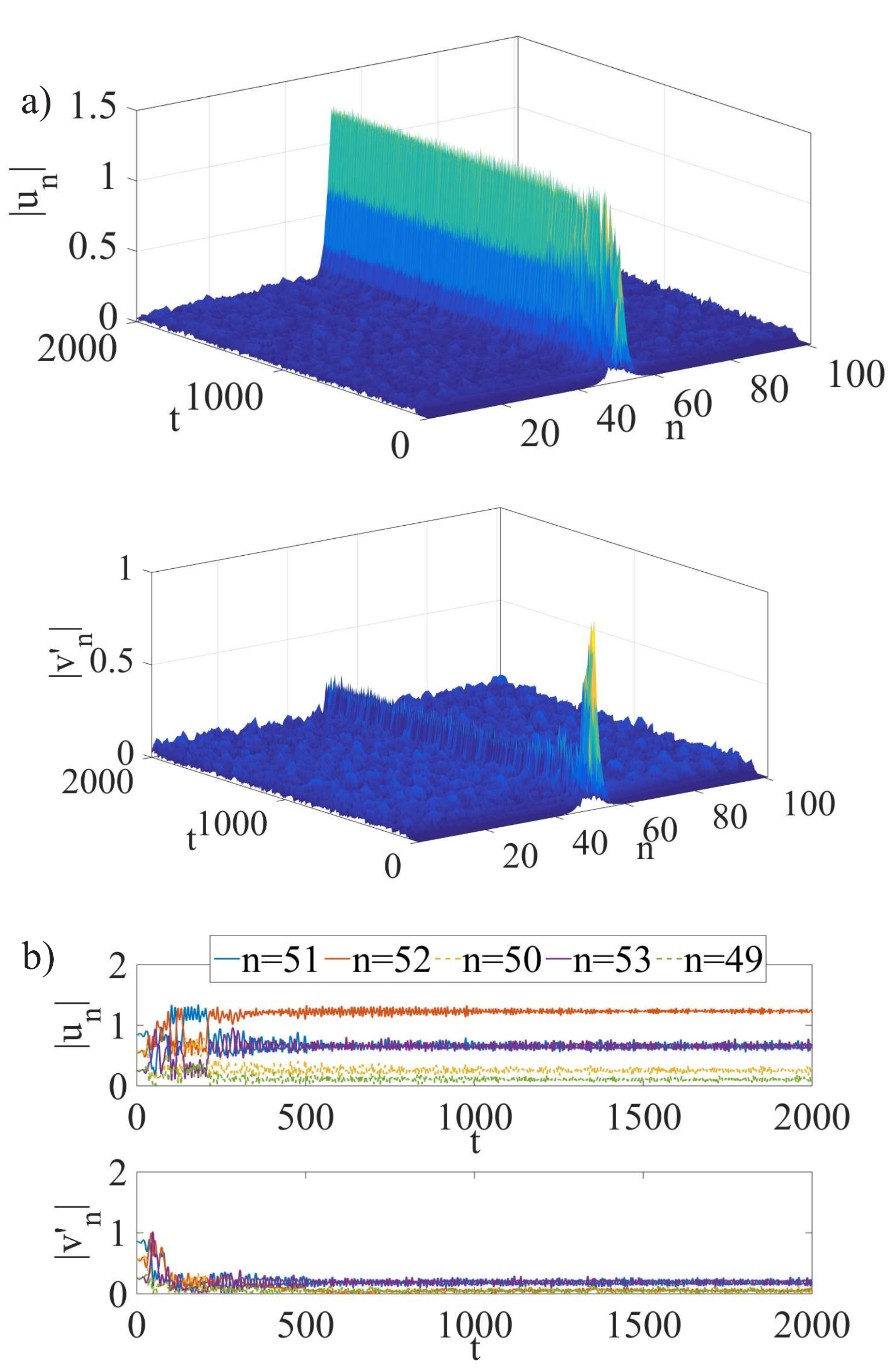}
\caption{Direct numerical simulation of a slightly randomly perturbed
spin-mixed mode with $\theta = 0$ and $\mu=2.5$, when
$\Gamma = 1$, $\sigma=0.5$, and $\mathfrak{a}=0.5$. Evolution of $u_{n}$ and ${v^{,}_{n}}$ components (a) and 
dynamics of corresponding components given specifically for the five central sites (b).}
\label{fig:unstab_mixed}
\end{figure}

An example of the dynamics that may result from the oscillatory instabilities
of the spin-mixed modes in this regime
is shown in Fig.\ \ref{fig:unstab_mixed}.
Note that, after the initial oscillatory dynamics, the solution settles down
at the stable fundamental spin-up mode (in this particular case the mode center
is also shifted one site to the right).

\begin{figure}
\includegraphics[width=0.33\textwidth]{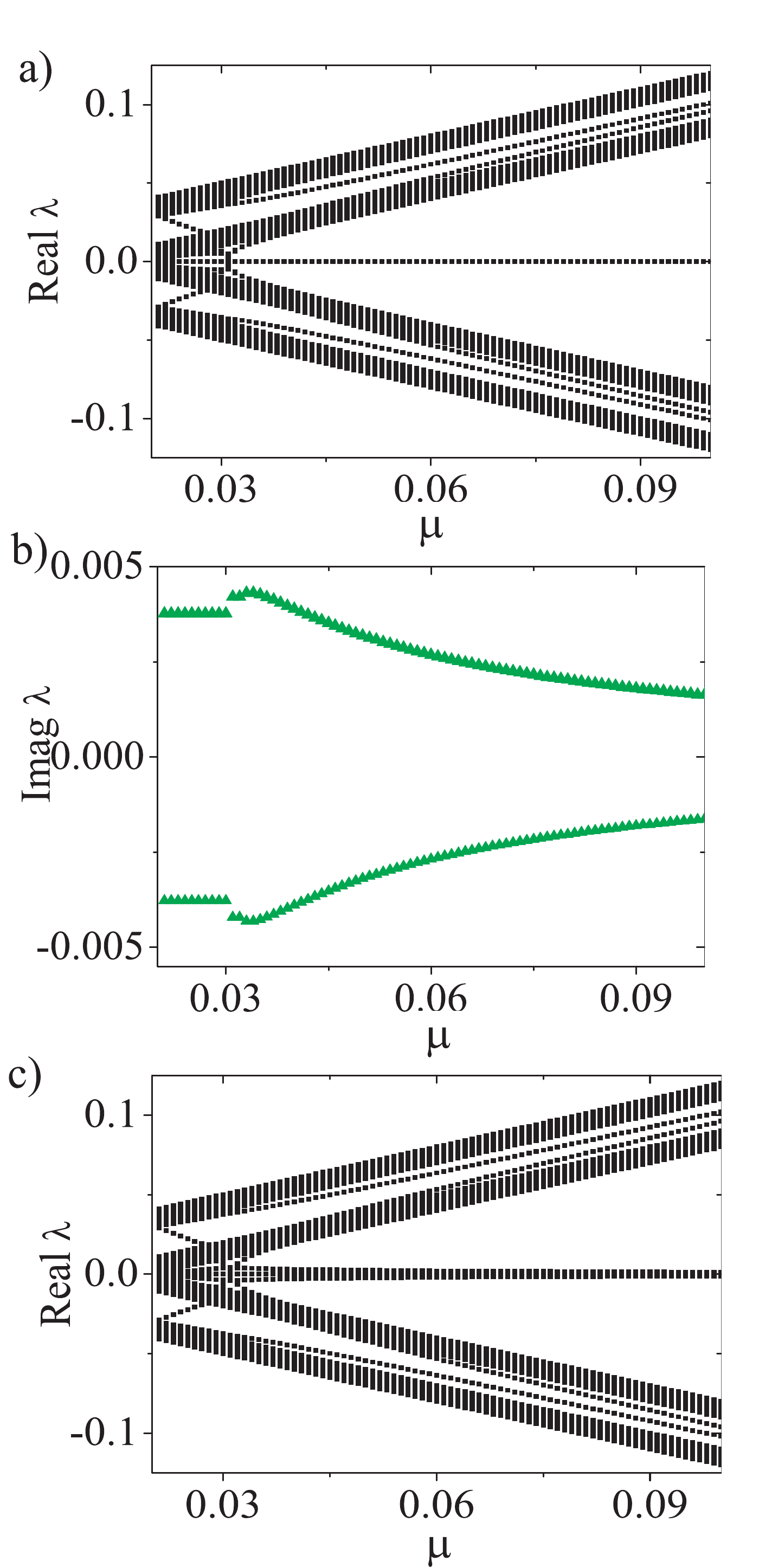}
\caption{Stability eigenvalues for the continuation of
fundamental spin-mixed states (\ref{eq:mixed}) with $\theta = \pi/2$ when
$\Gamma = 0.01$ and $\sigma=0.005$. Real (a) (Imaginary (b)) parts of
eigenvalues when  $\mathfrak{a}=0.5$. Eigenvalues when
$\mathfrak{a}=1.5$ (c) (imaginary parts are zero to numerical accuracy).
\label{fig:stab_mixedpi2}}
\end{figure}

As illustrated in Figs.\ \ref{fig:stab_mixed} (c,d),
the stability regime increases for $\mathfrak{a}$ increasing towards 1, and
exactly at $\mathfrak{a} = 1$ the spin-mixed states are always stable. However,
for $\mathfrak{a} > 1$ the eigenvalue pair originating from zero again goes out
along the imaginary axis (not shown in Fig.\ \ref{fig:stab_mixed})
as for $\mathfrak{a} < -1$, and thus spin-mixed modes with $\theta = 0$ are
generally unstable also for $\mathfrak{a} > 1$.
In fact, this latter instability can be considered as a stability exchange with
the $\theta = \pi/2$ spin-mixed mode, which, as illustrated in
Fig.\ \ref{fig:stab_mixedpi2},  is generally unstable with purely
imaginary eigenvalues for $\mathfrak{a} < 1$ (Figs.\ \ref{fig:stab_mixedpi2}(a,b)) but stable for $\mathfrak{a} > 1$ (Fig.\ \ref{fig:stab_mixedpi2}(c)).

\subsection{Compact modes in the flat-band limit}
\begin{figure}
\includegraphics[width=0.33\textwidth]{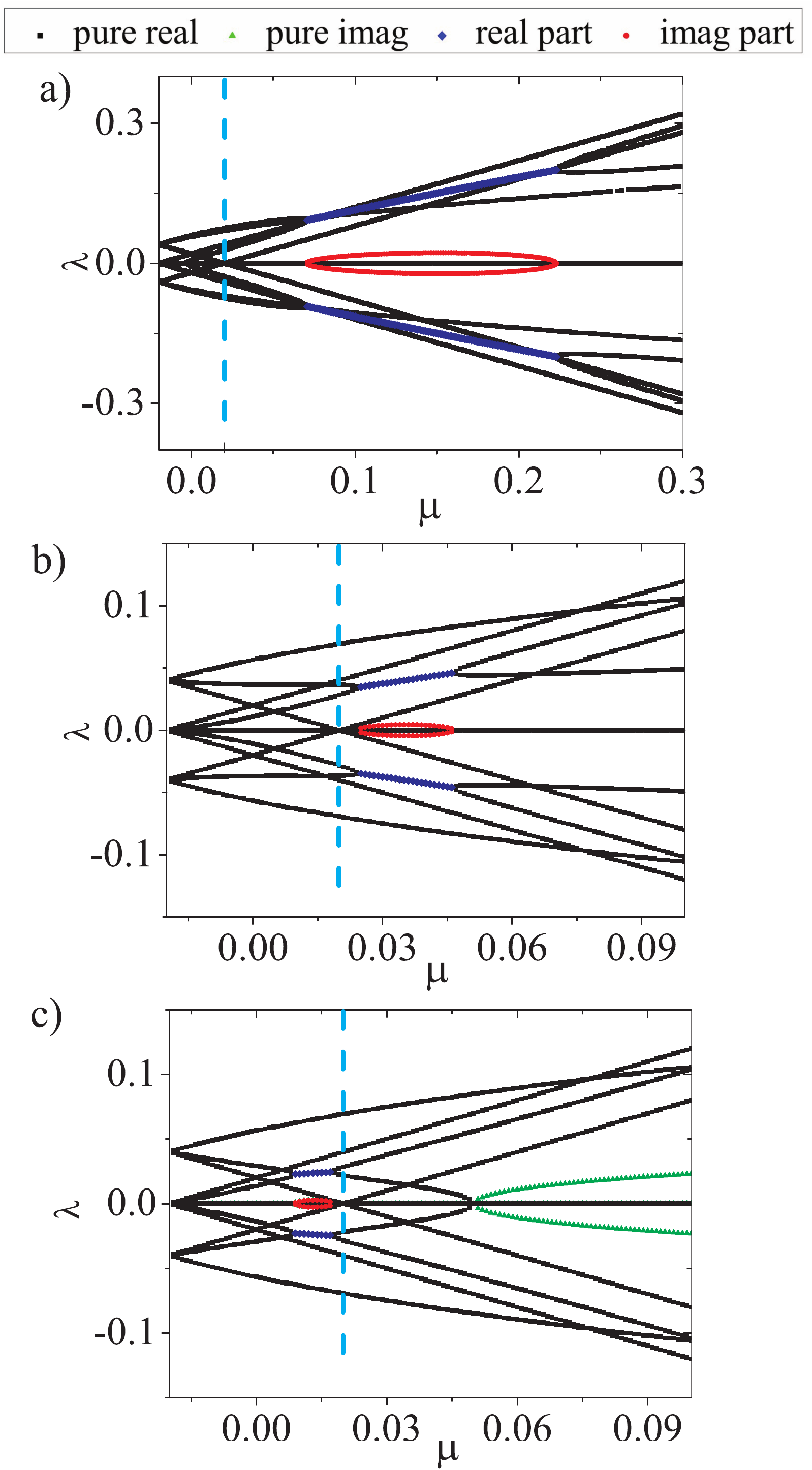}
\caption{Stability eigenvalues for the compacton (\ref{eq:compacton+})
with upper sign when
$\Gamma = \sigma=0.01$, and $\mathfrak{a}=-0.5$ (a),
 $\mathfrak{a}=0.5$ (b), and $\mathfrak{a}=1.5$ (c), respectively.
Purely real eigenvalues are
represented in black, while green (light gray) colored symbols stand for purely imaginary ones. 
Complex eigenvalues are represented in
blue (dark gray) and red (middle gray) for their real and imaginary parts,
respectively. The blue dashed vertical line represents the upper gap edge.}
\label{fig:stab_compact}
\end{figure}

In the flat-band limit, we may obtain exact analytical expressions for the
stability eigenvalues of the single two-site compacton modes.
We focus as above on the specific case with
$\Gamma = \sigma > 0$ and $\mathfrak{a}> -1$, when the nonlinear
compacton originating
from $\mu = -2 \Gamma$ (Eq.\ (\ref{eq:compacton+}) with upper sign) enters the
mini-gap for increasing $\mu$. For all zero-amplitude sites, the eigenvalues
are just those corresponding to the flat-band linear spectrum,
$\lambda = \pm \mu \pm 2 \Gamma$. For the compacton sites, four eigenvalues
correspond to local oscillations obtained by eliminating the surrounding
lattice: $\lambda = 0$ (doubly degenerate as always) and
$\lambda = \pm 2 \sqrt {2 \Gamma (\mu + 4 \Gamma)}$. Since the eigenvalues of
these internal modes are always real for $\Gamma > 0$ and they do not couple
to the rest of the lattice, they do not generate any instability. The remaining
eigenvalues describe the modes coupling the perturbed compacton to the
surrounding lattice, and are obtained from the subspace with
$c_{n_0} = i f_{n_0}, d_{n_0} = - i g_{n_0},
c_{n_0+1}= -if_{n_0+1}, d_{n_0+1}=ig_{n_0+1}$. The rather cumbersome result
can be expressed as:
\begin{eqnarray}
\lambda^2= \frac{\mu^2}{2} + 2\mu\Gamma\frac{1-\mathfrak{a}}{1+\mathfrak{a}}+
2\Gamma^2\left(1+\frac{4}{1+\mathfrak{a}}\right)\nonumber\\
\pm \left\{\frac{\mu^4}{4}
-\frac{2\Gamma\mu^3(1-\mathfrak{a})}{1+\mathfrak{a}}
+2\Gamma^2 \mu^2\left[1+4\frac{(1-\mathfrak{a})^2-2}{(1+\mathfrak{a})^2}
+\frac{4\mathfrak{a}}{1+\mathfrak{a}}\right]
+ \frac{8\Gamma^3\mu}{1+\mathfrak{a}}\left(1-\mathfrak{a}-\frac{8\mathfrak{a}}{1+\mathfrak{a}}\right)
+ 4\Gamma^4\left[\left(1+\frac{4}{1+\mathfrak{a}}\right)^2-4\right]
\right\}^{1/2} .
\label{eq:compactonev}
\end{eqnarray}
Oscillatory instabilities are generated if the expression inside the
square-root in (\ref{eq:compactonev}) becomes negative. Since obtaining
explicit general expressions for instability intervals in $\mu$ would require 
solving a nontrivial fourth-order equation, we show in
Fig.\ \ref{fig:stab_compact} numerical results
for the specific parameter values $\mathfrak{a} = \pm 0.5$ and $1.5$.
As can be seen, the compacton remains stable throughout the mini-gap as long
as $\mathfrak{a} \leq 1$ but
develops an interval of oscillatory instability in the semi-infinite gap above
the spectrum. The instability interval vanishes exactly at $\mathfrak{a} = 1$,
but then moves into the upper part of the mini-gap for $\mathfrak{a}>1$.
Purely imaginary eigenvalues, resulting from the terms outside the
square-root in (\ref{eq:compactonev}) becoming negative, also appear
in the semi-infinite gap for $\mathfrak{a}>1$.

\subsection{Gap modes in the mini-gap}
\begin{figure}
\includegraphics[width=10cm]{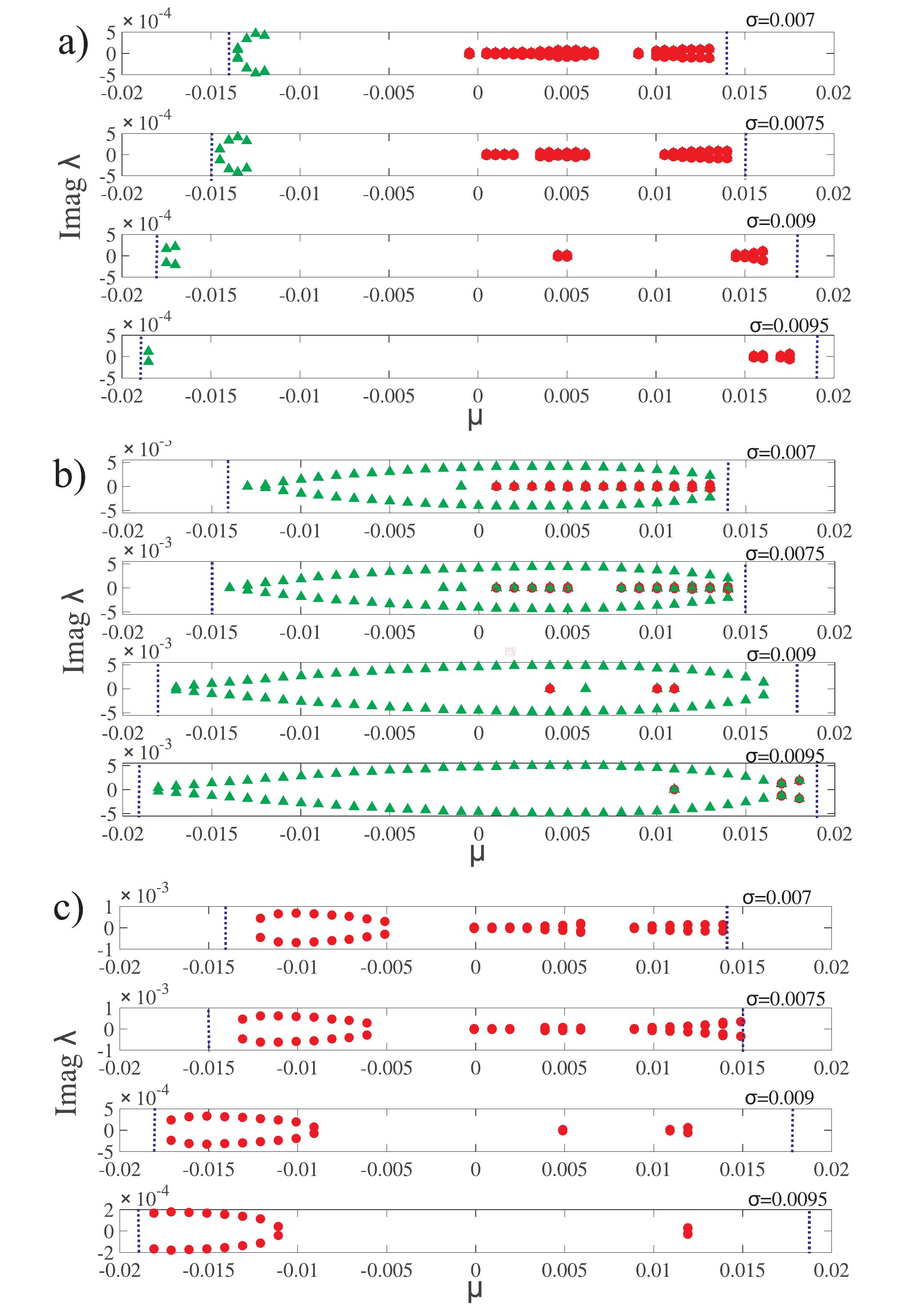}
\caption{Imaginary parts of stability eigenvalues for the continuation of: 
fundamental (I type) (a), 
type II (b) and type III (c) gap
mode inside the mini-gap, when $\Gamma = 0.01$ and
 $\mathfrak{a}=0.5$. Pure imaginary eigenvalues are depicted with green 
(light gray) triangles, 
while red (dark gray) circles correspond to imaginary parts of complex 
eigenvalues. 
When $\sigma=0.01$, the eigenvalues for fundamental gap mode are those of
the compacton illustrated in Fig.\ \ref{fig:stab_compact}(b). From
bottom to top, $\sigma$ is decreased to 0.007. Blue vertical dotted lines
represent the locations of the lower and upper gap edges. Profiles of the 
corresponding 
solutions at $\sigma=0.007$ in the mid-gap ($\mu=0$) are depicted in 
Fig.\ \ref{fig:gap_mode}.
}
\label{fig:stab_gap}
\end{figure}

\begin{figure}
\includegraphics[width=12cm]{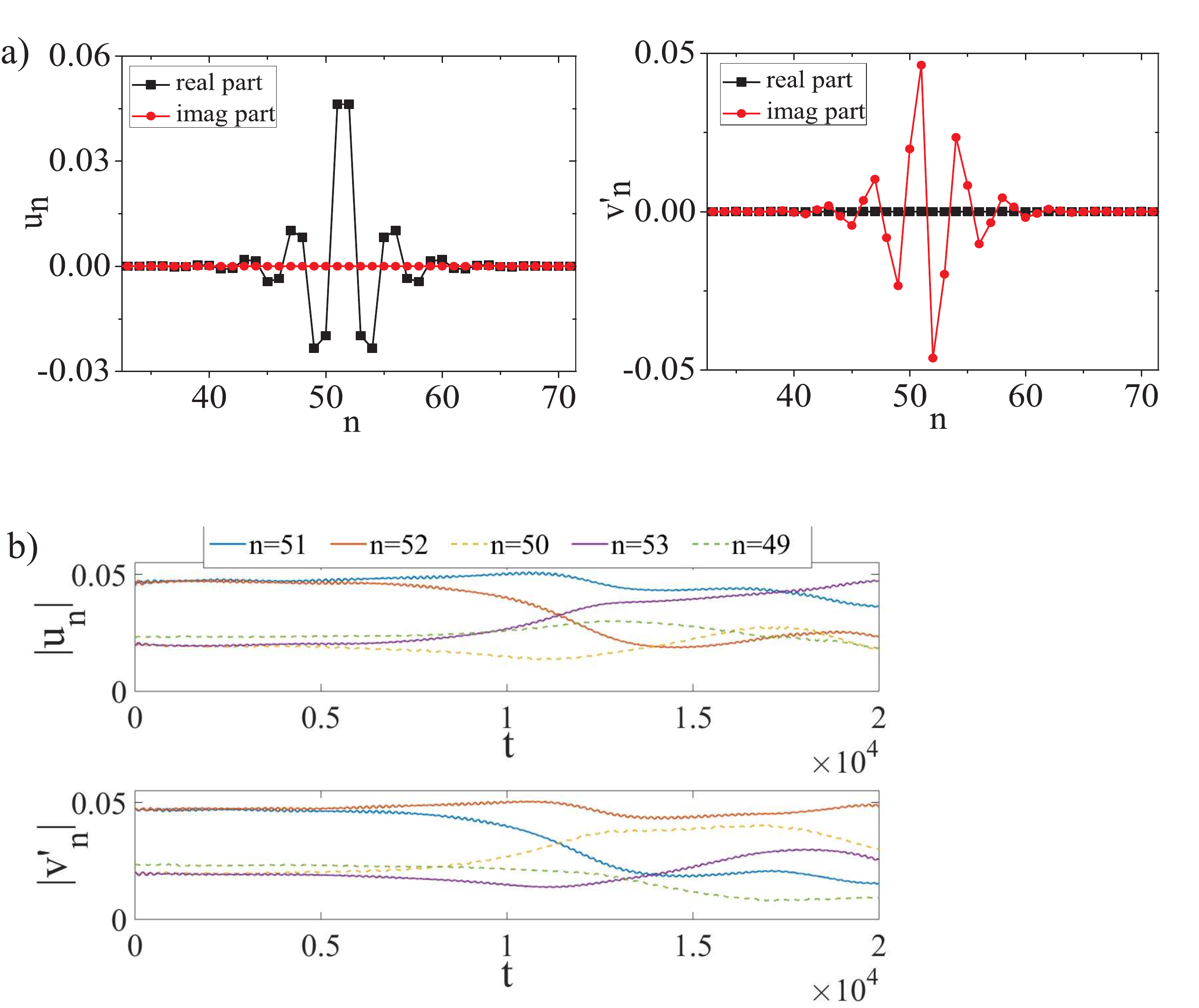}
\caption{Profile of an unstable fundamental gap mode
with $\Gamma = 0.01, \mathfrak{a}=0.5, \sigma = 0.007, \mu = -0.0125$ (a).
Direct simulation of the dynamics when this mode is slightly
perturbed; only five central sites are shown (b). Note the tendency for the
spin-up and spin-down components to localize mainly on odd and even sites,
respectively, after $t \sim 10^4$. }
\label{fig:unstab_gap}
\end{figure}

For the fundamental (type I) gap mode continued from the single two-site
compacton (\ref{eq:gap}) (assuming again  $\mathfrak{a} > -1$ and
$\Gamma > \sigma > 0$), we illustrate in Fig.\ \ref{fig:stab_gap}(a) typical
results for the numerical stability analysis.
As can be seen, as $\sigma$ decreases from the compacton limit
$\sigma = \Gamma$, weak instabilities start to develop mainly close to the
two gap edges. A further decrease in $\sigma$ yields instabilites in most of
the upper half of the gap, while the mode {\em remains stable in large parts of
the lower half}. 
Comparison with the stability eigenvalues for the exact
compacton (Fig.\ \ref{fig:stab_compact}(b)) shows that the instabilities
in the upper part of the gap result from resonances between modes
corresponding to compacton internal modes (\ref{eq:compactonev})
and the continuous linear spectrum modes, which get coupled as the
tail of the solution gets more extended. (These are seen in 
Fig.\ \ref{fig:stab_compact}(b) as eigenvalue collisions at 
$\mu \approx 0.005$ and  $\mu \approx 0.015$, but do not generate any 
instability in this figure since the corresponding eigenmodes are uncoupled 
in the exact compacton limit. However, they generate oscillatory instabilities 
when the exact compacton condition is not fulfilled, as seen in 
Fig.\ \ref{fig:stab_gap}(a).) 
On the other hand, the instabilities
appearing close to the lower gap edge, where the shape of the gap mode is
far from compacton-like and closer to a continuum gap soliton
(see Fig.\ \ref{fig:unstab_gap} (a))
arise from purely imaginary eigenvalues. Direct numerical simulations
of the dynamics in this regime (Fig.\ \ref{fig:unstab_gap} (b))
shows that the main outcome of these
instabilities
is a spatial separation of the spin-up and spin-down components.

As for the type II gap modes that arise in the mini-gap from the superposition 
of two 
in-phase neighboring single compactons in the presence of nonlinearity, we 
obtained  
pure imaginary eigenvalues in the whole mini-gap, even for the case when value 
of $\sigma$ slightly differs  
from $\Gamma$ (see Fig.\ \ref{fig:stab_gap} (b)). Here, with further decrease 
of $\sigma$, eigenvalues related to 
oscillatory instabilities start to occur but only in the upper half of the 
mini-gap. 

On the other hand, instability eigenvalue spectra for type III gap solutions 
contain only 
imaginary parts of complex eigenvalues (see Fig.\ \ref{fig:stab_gap} (c)).
These instabilities are always present in the lower half of the mini-gap and 
expand to the upper part as we move 
further from the compacton limit. 
   
\section{Conclusions}
\label{sec:conc}
We derived the relevant tight-binding model for a zigzag-shaped chain of 
spin-orbit coupled exciton-polariton condensates, focusing on the case with 
basis functions of zero angular momentum and chain angles $\pm\pi/4$. The 
simultaneous presence of spin-orbit coupling and nontrivial geometry opens 
up a gap in the linear dispersion relation, 
even in absence of external magnetic 
fields. At particular parameter values, where the strength of the 
dispersive and spin-orbit nearest-neighbor couplings are equal, the linear 
dispersion vanishes, leading to two
flat bands with associated compact modes localized at two neigboring sites.

We analyzed, numerically and analytically, the existence and stability 
properties of nonlinear localized modes, as well in the semi-infinite gaps  
as in the mini-gap of the linear spectrum. The stability of fundamental 
single-peaked modes in the semi-infinite gaps was found to depend 
critically on the parameter $\mathfrak{a}$ describing the
relative strength of the nonlinear interaction between polaritons of opposite
and identical spin (the latter assumed to be always repulsive). Generally, 
a spin-mixed mode with phase difference $\pi/2$ between spin-up and 
spin-down components is favoured when 
$\mathfrak{a}>1$ (cross-interactions repulsive and stronger than 
self-interactions), while a spin-polarized mode is favoured for 
$\mathfrak{a}<1$, which is the typical case in most physical setups. 
However, significant regimes of linear stability were found also for 
spin-mixed modes with zero phase difference between components when 
$|\mathfrak{a}|<1$, and for spin-polarized modes when $\mathfrak{a}>1$.

For parameter values yielding a flat linear band, nonlinear compactons appear 
in continuation of the linear compact modes, in the mini-gap as well as in the 
semi-infinite gaps. The linear stability eigenvalues for a single two-site 
compacton were obtained analytically, and shown to result in purely stable 
compactons inside the mini-gap when $\mathfrak{a}<1$, while regimes of 
instability were identified in the semi-infinite gaps, and 
when  $\mathfrak{a}>1$ also inside the mini-gap. Continuing compact two-site 
modes 
away from the exact flat-band limit yields the exponentially localized 
fundamental nonlinear gap modes 
inside the mini-gap. Several new regimes of instability develop, but the 
fundamental gap modes typically remain stable in large parts of the lower 
half of the 
gap   when $\mathfrak{a}<1$. We also found numerically 
nonlinear continuations of superpositions of two overlapping neighboring 
compactons (i.e., localized on three sites) with phase difference zero 
or $\pi/2$, where the latter also were found to exhibit significant regimes 
of linear stability in the mini-gap. 

The model studied here may have an experimental implementation with 
exciton-polaritons in microcavities.
Recently, microcavities have been actively investigated as quantum simulators 
of condensed matter systems.
Polaritons have been proposed to simulate XY Hamiltonian~\cite{Berloff-2017}, 
topological insulators~\cite{Nalitov-Z, Klembt18}, various types of 
lattices~\cite{Nalitov-graphene,Gulevich-kagome,Klembt17,Lieb-2018} 
among other interesting proposals~\cite{Sala-PRX-2015} many of which were
realized experimentally. In fact, the quasi one-dimensional zigzag chain 
considered here may be a more practical system to study the effects of 
interactions in presence of spin-orbit coupling as compared to the full 
two-dimensional systems mentioned above. A possible realization of the studied 
system could be using microcavity pillars or tunable open-access 
microcavities~\cite{Duff-APL-2014}. In the latter ones, large values of 
TE-TM splitting can be achieved exceeding that of monolithic cavities by a 
factor of three~\cite{Dufferwiel15}. Apart from directly controlling the 
strength of TE-TM splitting
by changing parameters of the experimental system such as the offset of the 
frequency from the center of the stop band of the distributed Bragg 
reflector~\cite{Panzarini-PRB-1999}, one more possibility to control 
parameters of the system is provided by using the excited states of the 
zigzag nodes such as spin vortices which were shown to influence the sign of 
the coupling strength between the sites in a polaritonic 
lattice~\cite{DY-graphene}. To what extent it is also possible to realize 
the exact tight-binding flat-band condition derived here, i.e., to tune 
experimental parameter values so that the nearest-neighbor spin-orbit coupling
coefficient $\sigma$ becomes equal to the standard dispersive
nearest-neighbor overlap integral $\Gamma$ while hoppings beyond 
nearest neighbors remain negligible, is to the best of our 
knowledge an open question. 

Finally, we note also the recent realizations of
zigzag chains with large tunability for atomic Bose-Einstein
condensates~\cite{An18}, opening up the possibility for studying related
phenomena
involving spin-orbit coupling in a different context. Having in mind 
experimental progress on coherent transfer of atomic Bose-Einstein condensates 
into the flat bands 
originating from different optical lattice configurations 
(e.g., \cite{Taie,Kang}), as well as in engineering spin-orbit coupling 
within ultracold atomic systems \cite{Lin,Zhai}, experimental realization 
of phenomena analogous to those described in the present work should be 
expected to be within reach. Very recently, a theoretical proposal for 
observing flat bands and compact modes for spin-orbit coupled atomic 
Bose-Einstein condensates in one-dimensional shaking optical lattices also 
appeared, where an exact tuning of the spin-orbit term could be achieved by 
an additional time-periodic modulation of the Zeeman field \cite{AS18}.

\begin{acknowledgments} We thank Aleksandra Maluckov and Dmitry Yudin
 for useful discussions.
We acknowledge support from the European Commission
H2020 EU project 691011 SOLIRING, and from the Swedish Research Council
through the project ``Control
 of light and matter waves propagation and localization in photonic lattices''
within the Swedish Research Links programme,
 348-2013-6752. P.P.\ Beli\v cev and G.\ Gligori\'c acknowledge support from 
the Ministry of Education, Science and Technological Development of
Republic of Serbia (project III45010).

\end{acknowledgments}

\end{document}